# Effect of annealing on the corrosion-fatigue strength and hot salt corrosion resistance of fine-grained titanium near-α alloy Ti-5Al-2V obtained by Rotary Swaging

V.N. Chuvil'deev[a], A.A. Murashov[a], A.V. Nokhrin[a,(*)], N.N. Berendeev[a], C.V. Likhnitskii [a], A.N. Sysoev[a], N.V. Melekhin[a], K.A. Rubtsova[a], A.M. Bakmetyev[b], P.V. Tryaev[b], R.A. Vlasov[b], N.Yu. Tabachkova[c, d], A.I. Malkin[e]

[a] Lobachevsky State University of Nizhny Novgorod, 603022 Russia, Nizhny Novgorod, Gagarina ave., 23

[b] Afrikantov Experimental Design Bureau for Mechanical Engineering JSC (Afrikantov OKBM JSC), 603074 Russia, Nizhny Novgorod, Burnakovsky proezd, 15

[c] National University of Science and Technology "MISIS", 119991 Russia, Moscow, Leninskiy ave, 4

[d] A.M. Prokhorov General Physics Institute of the Russian Academy of Science, 119991 Russian, Moscow, Vavilova st., 38

[e] A.N. Frumkin Institute of Physical Chemistry and Electrochemistry of the Russian Academy of Science, 119071 Russia, Moscow, Leninskiy ave., 31

e-mail: nokhrin@nifti.unn.ru

**Abstract**

The corrosion-fatigue strength in 3% aqueous NaCl solution and the resistance against hot salt corrosion (HSC) of the fine-grained near-α alloy Ti-5Al-2V (Russian analog of Grade 9 titanium alloy with increased aluminum content) has been studied. The properties of the Ti-5Al-2V alloy in the coarse-grained state, in the fine-grained after cold Rotary Swaging (RS), in partly recrystallized state, and in fully recrystallized one have been investigated. The mechanical properties of the alloy were characterized using compression tests and microhardness measurements. The effects of RS and of the annealing temperature and time on the character of corrosion destruction of the surface and on the composition of the products of the HSC were studied. RS was shown to result in an increase in the depth of the intergranular corrosion defects while the recrystallization annealing promotes the increasing of the corrosion resistance of the Ti-5Al-2V titanium alloy. The parameters of the Basquin equation for the corrosion-fatigue curves for the near-α Ti-5Al-2V alloy in the coarse-grained state, in the severely strained one, and after recrystallization annealing were determined for the first time. An effect of nonmonotonous dependencies of the slopes of the corrosion-fatigue curves for the strained near-α Ti-5Al-2V alloy on the recrystallization annealing temperature has been observed.

---

(*) Corresponding author (nokhrin@nifti.unn.ru)

**Highlights**

- The hat salt corrosion (HSC) resistance of the near-α Ti-5Al-2V alloy depends on the presence of β-phase particles and the concentration of vanadium at grain boundaries
- Cold Rotary Swaging (RS) leads to an increase in the strength, hardness, and corrosion-fatigue strength of the Ti-5Al-2V alloy
- RS results in a reduction in the corrosion resistance of the titanium alloy Ti-5Al-2V
- Recrystallization annealing leads to an improvement in the resistance to intergranular HSC
- Recrystallization annealing leads to a non-monotonic change in the slope of the corrosion fatigue curve $\sigma_a(N)$

**1. Introduction**

Titanium α- and near-α alloys are used widely for manufacturing high-strength corrosion-resistant structures in nuclear power engineering designed for long-term operation under corrosive environment conditions, elevated temperatures, static and cycled stresses, and, in some cases, radiation exposure [1-3]. Modern titanium α- and near-α alloys are required to exhibit simultaneously increased strength, corrosion resistance in operational conditions, and corrosion-fatigue strength. This necessitates extensive investigations of the mechanisms of hot salt corrosion (HSC), stress corrosion cracking (SCC), and corrosion fatigue of titanium alloys [1-5].

At present, the mechanisms of HSC, hot salt SCC and fatigue failure of coarse-grained titanium alloys have been studied thoroughly. The mechanisms of these damaging processes in titanium alloys are described in detail in many classical works (see, for example, [6-13] etc.) and we will not analyze them in detail here. The effects of grain size and microstructures type (see [7, 14-17], etc.), chemical compositions of titanium alloys [18-21], surface treatments [22-26], test conditions [6, 12, 13, 27-29], etc. have been investigated. The effect of HSC on the fatigue characteristics of titanium alloys has been studied [30, 31], the effect of hydrogen on resistance to hot salt SCC of titanium alloys [16, 32], the importance of oxygen accounting in the processes of hot salt SCC of titanium alloys [33], etc. has been studied. In recent years, the mechanisms of hot corrosion and corrosion fatigue failure in fine-grained and ultrafine-grained (UFG) high-strength titanium alloys were studied [34-37]. The high resistance of UFG Ti-Al-V(Zr) α- and near-α alloys to intergranular HSC was shown to originate from a reduction in the local concentration of corrosive elements at grain boundaries (primarily vanadium) and the refinement of β-phase particles [35, 36].

The interest to UFG titanium alloys is driven primarily by their promising applications in mechanical engineering. UFG titanium alloys exhibit high strength [35-38], high fatigue resistance [37-39], sometimes simultaneously increased strength and ductility [38, 40], high corrosion resistance [36, 41, 42] and SCC resistance [43], etc. The use of high-strength UFG titanium alloys while maintaining their high corrosion resistance allows manufacturing thinner and lighter structural elements. High ductility of UFG titanium alloys at elevated temperatures [44, 45] enables the production of such thin and light structural elements using superplastic stamping technology. It should be noted that UFG titanium α- and near-α alloys have low thermal stability; grain growth may occur upon heating [35, 46]. The grain growth process leads to a decrease in strength and many operational characteristics of UFG titanium alloys. The low temperature of grain growth initiation limits the permissible temperatures and service times for structures and products that require high strength and reliability. The low thermal stability of the UFG microstructure of titanium alloys necessitates the use of new technologies such as high-speed solid-phase diffusion welding to preserve the UFG microstructure in welded joints [47, 48].

The goal of the present work was to study the effect of annealing on the corrosion resistance and corrosion-fatigue strength of the UFG near-α titanium alloy Ti-5Al-2V. The UFG microstructure in the alloy was formed by Rotary Swaging (RS). RS technology is an effective way to manufacture functionally-graded materials in which the surface layer has a smaller grain size and a higher hardness than the center of the sample [49-52]. This makes it possible to effectively combine high strength and plasticity in UFG samples made by the RS method. It was previously shown that RS allows to increase the corrosion resistance and fatigue strength of Ti-2.5Al-2.6Zr titanium α-alloy [36, 37]. Currently, the materials obtained by the RS method are considered as functional gradient materials, the use of which is promising in mechanical engineering and energy [49, 50, 52].

No studies of the effect of annealing on the susceptibility of the UFG Ti-5Al-2V alloy to HSC and corrosion-fatigue failure have been reported to date. By choosing the temperature of a 30-minute annealing, we varied the volumetric fraction of recrystallized structure and the average grain size. By varying the holding time at a temperature of 250 °C, an assessment was made of the thermal stability of the non-equilibrium microstructure of the UFG Ti-5Al-2V alloy for the resistance to HSC of products made from titanium alloys.

## 2. Materials and Methods

The object of the study was the Russian industrial near-α titanium alloy grade PT-3V with a composition of Ti-5wt.%Al-2wt.%V, which is a Russian equivalent of titanium alloy Grade 9 Ti-(2.5-3.5)Al-(2.0-3.0)V with increased aluminum content. This titanium alloy is widely used in nuclear power and has high radiation resistance [3, 53]. The composition of the alloy conforms to the requirements of the Russian National Standard GOST 19807–91. The initial materials were rods of the PT-3V alloy with a diameter of 20 mm, manufactured at the Chepetsky Mechanical Plant (Glazov, Russia).

The UFG microstructure in the titanium alloy Ti-5Al-2V was formed by RS at room temperature on the rotary swaging machine RS5-4-21 HIP. The bars were deformed using four hammers made of high-strength steel. Rotary swaging of titanium rods from an initial diameter of 20 mm to a diameter of 12 mm was carried out in 2 mm increments (∅20 mm → 18 mm → 16 mm → 14 mm → 12 mm), next, in 1 mm steps (∅12 mm → 11 mm → … → 7 mm → 6 mm). The total accumulated strain during RS was 70%. The average strain rate was 0.6-1 $s^{-1}$. The rotation speed of the workpiece is ~ 60 rpm, the strike frequency of the strikers is 1200 $min^{-1}$. The heating temperature of the outer surface of titanium rods at RS did not exceed 60-100 °C.

The samples were subjected to annealing in two different regimes. Regime I involved a 30-minute annealing of titanium samples in the air at 500, 550, 600, 650, and 700 °C. Regime II involved annealing at 250 °C for durations of 500, 1000, 1500, 2000 h. The samples were cooled in air. Annealing under Regime I was used to study the evolution of the microstructure in the UFG titanium alloy at various temperatures. Annealing under Regime II was intended to test the long-term thermal stability of the UFG structure of the alloy.

Microstructure analysis was carried out using scanning electron microscopes (SEMs) Tescan Vega 2 and Jeol JSM-6490 with an EDS microanalyzer Oxford Instruments INCA 350 as well as with a transmission electron microscope (TEM) Jeol JEM-2100F.

X-ray diffraction (XRD) phase analysis of corrosion products was performed using the XRD-7000 diffractometer (CuKα radiation, scanning step 0.04°, exposure time at each step 2 s, scanning range of angles 2θ = 30 – 80°). Analysis of corrosion products of titanium alloy after testing for HSC was not possible due to the high content of the NaCl phase, which manifested very high intensities of XRD peaks. Therefore, before the XRD measurements, the corrosion products were collected mechanically from the surfaces of the samples, then washed with hot distilled water through filter paper. The resulting insoluble residue was dried in air (for at least 10 min) and ground in an agate mortar to obtain a homogeneous powder. Despite the steps undertaken, not the whole NaCl phase was removed from the corrosion products of the titanium alloys. As a result, highly intensive XRD peaks of the NaCl phase were present in the XRD curves of most samples examined. However, the significant reduction in their intensities as compared to the original samples allows for a qualitative analysis of their phase composition. The analysis of the results was carried out using the databases PDF-2 (ICDD, release 2012), PDF-4 (ICDD, release 2014), and ICSD (2016).

Microhardness (Hv) measurements were performed using the Qness 60A+ hardness tester. Compression tests of cylindrical specimens with a diameter of 5 mm and a height of 15 mm were carried out using the 2167 P-50 machine. During the compression tests by GOST 25.503-97, the stress-strain curves were obtained, which the yield strength of the titanium alloy ($\sigma_{0.2}$) was determined from. Stress-relaxation tests were performed according to the technique described in [54] by GOST R 57173-2016. Rectangular specimens of ∅5 mm in cross-section and of 10 mm in height were made for tests. Compression test samples were cut from the central part of a 6 mm diameter rod. As a result, a dependence of stress relaxation ($\Delta\sigma_i$) on summary stress ($\sigma$) was obtained. The curve was also used to determine lattice friction stress (macroelasticity stress) ($\sigma_0$) and yield strength ($\sigma_y$). In accordance with [54], the value of the macroelasticity stress corresponds to the threshold stress of the crystal lattice flow (lattice friction stress) of the material.

Corrosion-fatigue tests were conducted using the rotating bending scheme in a 3% aqueous solution of NaCl. Smooth cylindrical specimens with a diameter of 5 mm and a working part diameter of 3 mm (Type II according to GOST 25.502–79) were used. The samples were cut from the central part of the rod. The stress cycle asymmetry coefficient was $R_\sigma = -1$. The surface roughness of the working part in the specimens was Rz = 3.2 μm. The experiment involved recording the dependence of the number of cycles to failure (N) on the amplitude of stress applied ($\sigma_a$). The stress amplitude was calculated using solid mechanics methods. The average uncertainty in determining $\sigma_a$ was 10 MPa. Fatigue curves were analyzed using the power equation of Basquin $\sigma_a = AN^{-q}$, where $A$ and $q$ are numerical coefficients [55]. Fractographic analysis of fracture surfaces of the specimens was conducted using SEM.

The autoclave tests for HSC were conducted in NaCl salt at a temperature of 250°C, in the presence of oxygen. The tests were carried out for 250 h in laboratory autoclaves with a volume of 3000 cm$^3$. For the tests, the samples were placed in the center of ceramic containers with an internal volume of 600 cm$^3$, which provided access to air and were filled with crystalline salt. Ceramic containers with samples of titanium alloys immersed in salt were placed in the centers of the test autoclaves. The temperature was maintained with an accuracy of ± 5°C. The testing procedure was described in details in [56]. The cylindrical samples with a diameter of 5 mm and a length of 15 mm were used for the tests. The sample surfaces were subjected to mechanical processing before testing. To determine the depth of corrosion, a cross-section of the sample was made. Corrosion products were removed by chemical methods following the requirements of GOST R 9.907-2007. The corrosion depth was determined using the Leica IM DRM optical microscope according to GOST 9.908-85. The average uncertainty of determining the depth of the corrosion defect was ±10 μm.

## 3. Results

### 3.1 Microstructure and Mechanical Properties

The microstructure of the coarse-grained Ti-5Al-2V alloy in its initial state was described in detail in [35], and we will not dwell on this issue in details here. It is important to stress that the microstructure of the titanium alloy in the initial state is highly nonuniform – the microstructure of the alloy consists of equiaxial α-grains up to 50 μm in size surrounded by areas of smaller plate-like α'-grains (Fig. 1a, b), along the boundaries of which precipitation of β-phase particles can be observed (Fig. 1b). Such a microstructure is conventional for titanium near- α alloys obtained by hot deformation with a gradual decrease in temperature from the β-region to the (α+β)-region [57]. The microhardness of the coarse-grained alloy in the initial state is Hv = 1.9-2.0 MPa. The yield strength of a titanium alloy in its initial state is $\sigma_{0.2}$ ~ 735-800 MPa, which is 135-150 MPa higher than the minimum allowable yield strength for this alloy ($\sigma_{0.2} \geq 590$-$600$ MPa). The increased strength of the Ti-5Al-2V alloy is associated with the presence of β-phase particles along grain boundaries (Fig. 1b), which prevent the movement of dislocations through grain boundaries. The value of the Hall-Petch coefficient $K = (\sigma_y - \sigma_0)\sqrt{d}$ of a coarse-grained titanium alloy varies from 1.9 MPa·m$^{1/2}$ (at $d = 20$ μm) to 3.0 MPa·m$^{1/2}$ (at $d = 50$ μm).

An inhomogeneous macrostructure is formed in the titanium rod after the RS. The vortex-like macrostructure characteristic of the RS is clearly visible on the cross section of the rod (Fig. 1c, see [49-52]). After RS, a UFG microstructure with an average fragment size of ~0.2-0.5 μm was formed in the alloy. (Fig. 1d). Analysis of the results of electron microscopy has shown that the fine-grained microstructure obtained by RS can be characterized as a mixed grain-subgrain microstructure (Fig. 1d). No large β-phase particles were found in the microstructure of UFG alloys. Probably their strong grinding occurred during RS. The average microhardness of the UFG alloy was 3.0-3.1 GPa. The hardness of the samples surface layer was 150-200 MPa higher than the one of the central part of the sample (Fig. 2). The yield strength $\sigma_{0.2}$ of the alloy increases to 795 MPa after RS, and the physical yield strength $\sigma_y$ increases to 1070 MPa (Table 1). After RS, the value of the Hall-Patch coefficient of Ti-5Al-2V titanium alloy decreases to K ~ 0.45-0.5 MPa·m$^{1/2}$ (at $d = 0.5$ μm).

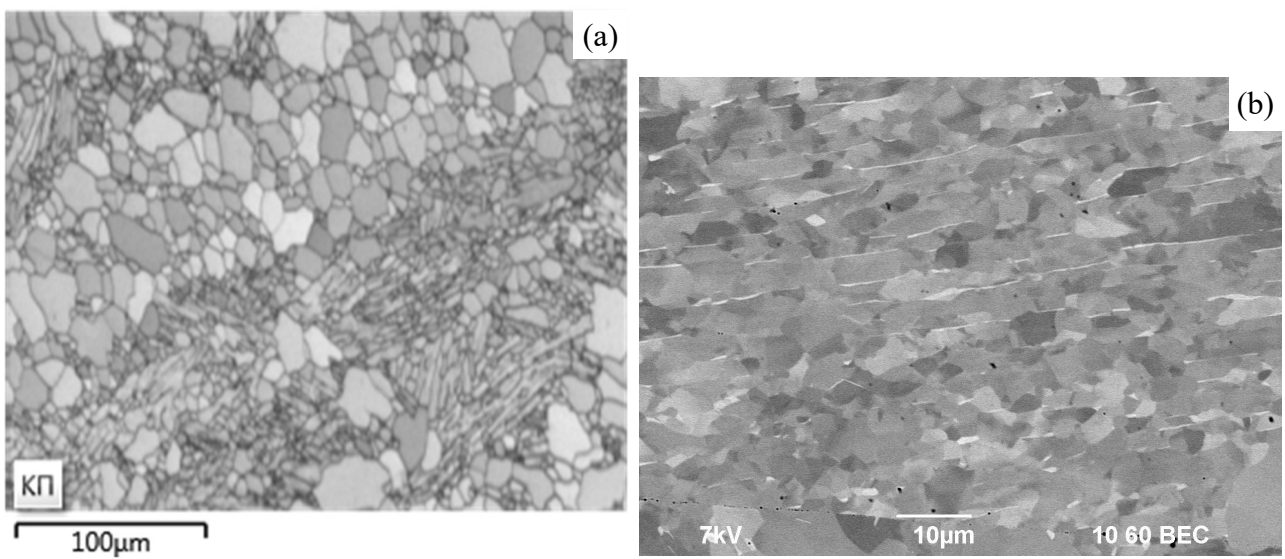

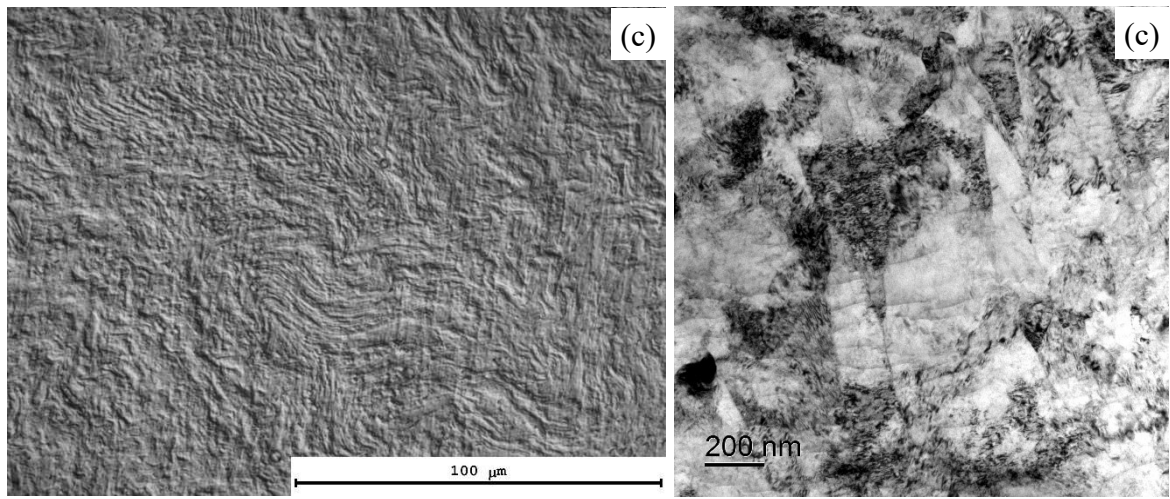

**Fig. 1** Structure of the Ti-5Al-2V alloy: (a, b) microstructure of the coarse-grained alloy (SEM) [35]; (c) macrostructure of the UFG alloy; (d) microstructure of the UFG state after RS (TEM) [35]

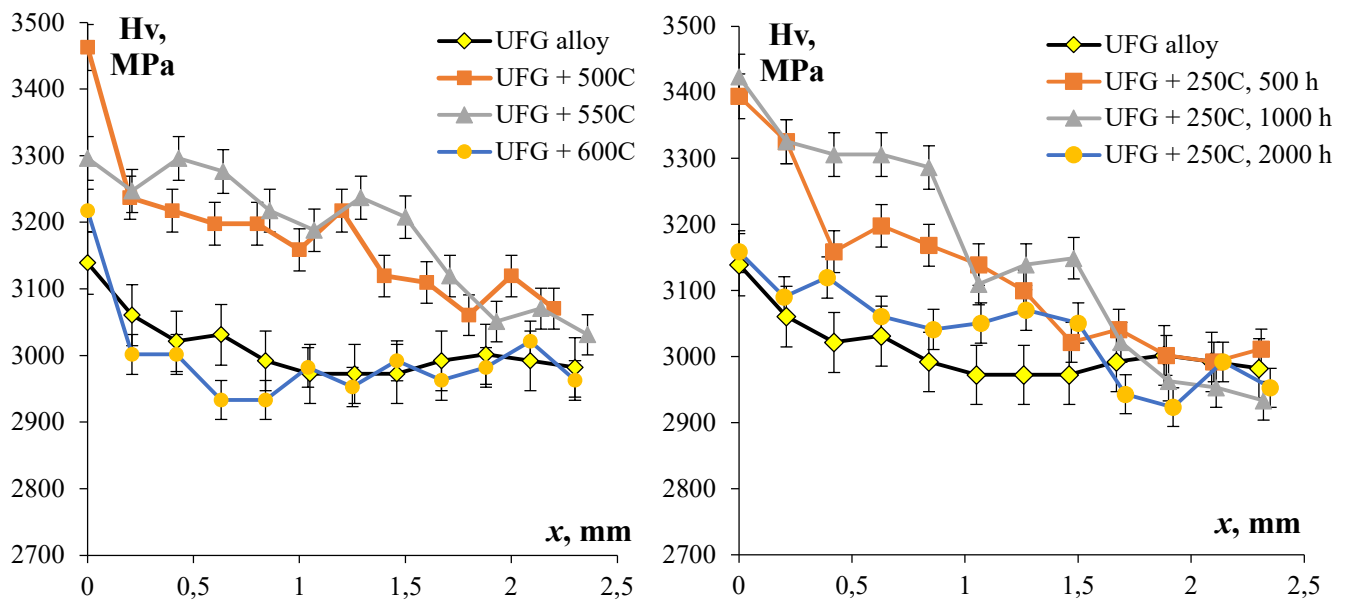

**Fig. 2** Distribution of microhardness along the bar diameter after different treatment modes: (a) the effect of the 30-minute annealing temperature; (b) the effect of the annealing time at 250 °C

The results of metallographic and electron microscopy studies show that prolonged isothermal annealing at a temperature of 250°C does not lead to any change in the microstructure of the UFG titanium alloy. After heating to a temperature of 550-600°C, some individual recrystallized grains with a size of ~0.5-1 μm were observed against a background of the severely deformed microstructure. Thus, after annealing at temperatures of 550-600°C, a partially recrystallized microstructure was formed in the UFG titanium alloy Ti-5Al-2V. Therefore, the recrystallization temperature in the UFG Ti-5Al-2V alloy after RS is close to the grain growth start temperature in the UFG Ti-5Al-2V alloy obtained by ECAP [35]. After a 30-minute annealing at a temperature of 700°C, a fully recrystallized microstructure with grain sizes ranging from 1 to 10 µm was observed (Fig. 3). The central parts of the samples had a larger grain sizes than the edge ones (Fig. 3). This is due to the nonuniform distribution of strain from the sample surface to the center during RS.

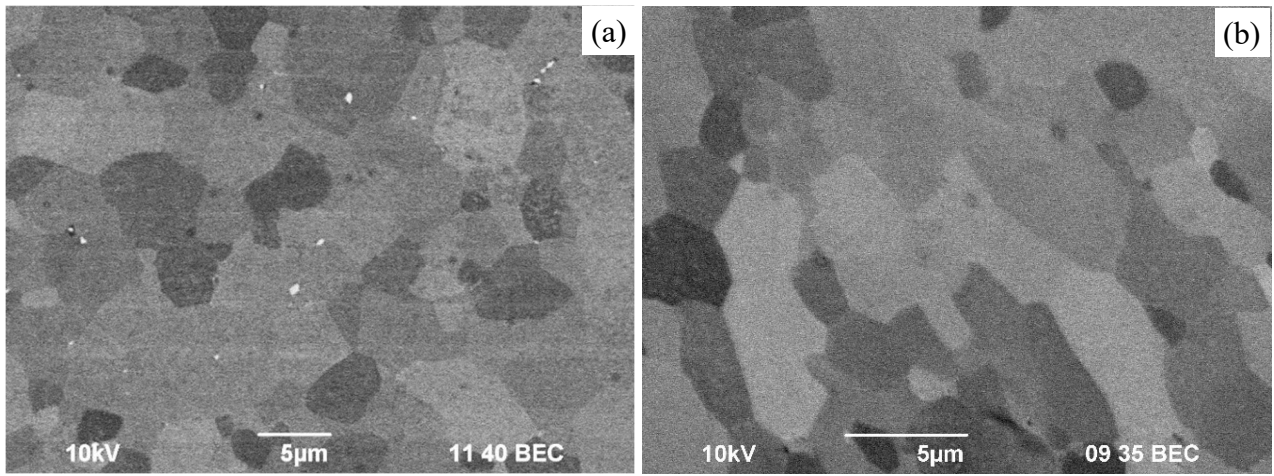

**Fig. 3** Microstructure of the Ti-5Al-2V alloy after annealing at 700°C for 30 minutes: (a) edge of the sample; (b) center of the sample. SEM

The results of the microhardness investigation of the annealed alloys are presented in Fig. 2. From Fig.2a, one can see that the microhardness of the Ti-5Al-2V alloy increases when heated up to 550°C and exceeds the microhardness of the unannealed UFG alloy. The microhardness of the UFG alloy also increases after 500-1000 hours of exposure at 250 °C (Fig. 2b). It should be noted that the uneven distribution of microhardness over the diameter of the titanium rod remains (Fig. 2). The effect of increasing microhardness in the Ti-5Al-2V UMP alloy is associated with the release of TiC nanoparticles [35]. After heating up to 600-700°C, a decrease in microhardness was observed attributed to the recovery and to the start of grain growth (Fig. 3). The unevenness of the hardness distribution in annealed alloys decreases, primarily due to a decrease in the microhardness of the surface layer (Fig. 2).

Table 1. Results of measurement of mechanical properties and corrosion-fatigue tests of the UFG Ti-5Al-2V alloy after high-temperature annealing for 30 min

| T, °C | CG alloy | UFG alloy | Annealing temperature, °C | | | |
|---|---|---|---|---|---|---|
| | | | 500 | 550 | 600 | 700 |
| $\sigma_0$, MPa | 370 ± 20 | 415 ± 30 | 450 ± 20 | 500 ± 25 | 530 ± 25 | 320 ± 15 |
| $\sigma_y$, MPa | 800 ± 40 | 1070 ± 50 | 1030 ± 50 | 1040 ± 50 | 870 ± 45 | 640 ± 30 |
| $\sigma_{0.2}$, MPa | 735 ± 10 | 795 ± 25 | 940 ± 30 | 985 ± 30 | 880 ± 30 | 665 ± 20 |
| A, MPa | 1166-1717 | 735 | 2021 | 2779 | 10321 | 1822 |
| Q | 0.09-0.011 | 0.02 | 0.12 ± 0.02 | 0.16 ± 0.02 | 0.31 ± 0.06 | 0.13 ± 0.02 |

The results of the comparative tests of the samples are summarized in Table 1. The stress-strain curves of the samples are shown in Fig. 4. The analysis of the results shows that RS leads to an increase in the yield strength as well as an increase in the flow stress corresponding to the onset of the stage of uniform plastic deformation of the UFG material. Annealing results in a reduction in the yield strength and flow stress of the UFG material. It is noteworthy that after a 30-min annealing at 700°C, the mechanical properties of the UFG alloy decrease to values that are lower than the one of the coarse-grained alloy in its initial state (Table 1, Fig. 4). This is quite an unexpected result, as it is assumed traditionally that annealing of UFG metals leads to a decrease in mechanical properties to the level corresponding to the performance of the coarse-grained metal.

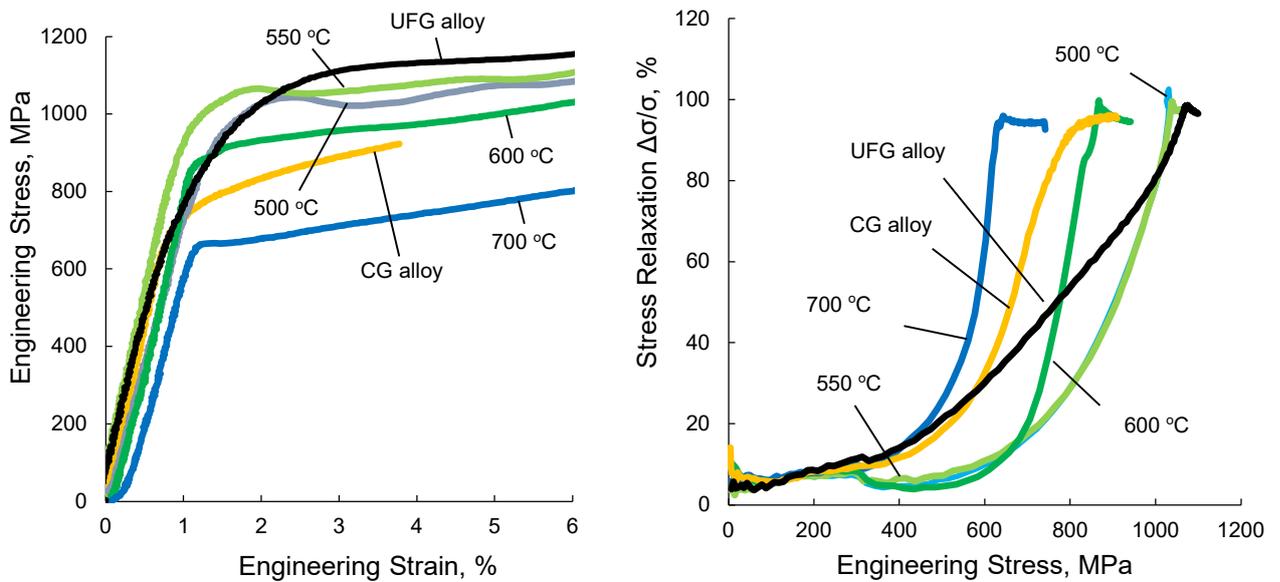

**Fig. 4** Stress-strain curves for the titanium alloy Ti-5Al-2V samples during compression tests: (a) tests according to GOST 25.503-97; (b) stress-relaxation tests according to GOST R 57173-2016

Attention should be paid to the results of stress-relaxation compression tests shown in Fig. 4b and in Table 1. From the Table 1 it can be seen that the dependence $\sigma_0(T)$ has a maximum at 550-600 °C. Previously, a similar dependence of $\sigma_0(T)$ was found for the UMP alloy Ti-5Al-2V obtained by the ECAP method and using *in-situ* TEM it was shown that it is due to the release of titanium carbide particles [35].

### 3.2 Corrosion-fatigue tests
#### *3.2.1 Effect of Rotary Swaging*

The results of the corrosion-fatigue tests of the coarse-grained and UFG Ti-5Al-2V alloy samples in the unannealed state are presented in Fig. 5. The slope of the $\sigma_a(N)$ curve for the UFG alloy is much greater than the one for the coarse-grained alloy. Note also that the UFG alloy has much higher cyclic durability – at a stress amplitude of 590-600 MPa, the number of cycles to failure for the coarse-grained alloy is $(0.9-1.5) \times 10^4$ cycles while for the UFG alloy it exceeds $10^5$ cycles. The fatigue limit for the UFG alloy based on $N^* = 3 \cdot 10^6$ cycles is $\sigma_{-1} = 600-610$ MPa, which significantly exceeds the fatigue limit of the coarse-grained alloy ($\sigma_{-1} = 300-320$ MPa). The relative fatigue limit of the near-α Ti-5Al-2V UFG alloy is close to the fatigue limit of the Ti-2.5Al-2.6Zr α-alloy obtained by the RS method ($\sigma_{-1}$ = 570-590 MPa based on tests $N^*$ = 3·$10^6$ cycles [37]). Analysis of the $\sigma_a(N)$ curves using the Basquin equation shows that the formation UFG microstructure be the RS method leads to a decrease in the coefficient A from 1166-1717 to 735 MPa (see Table 1).

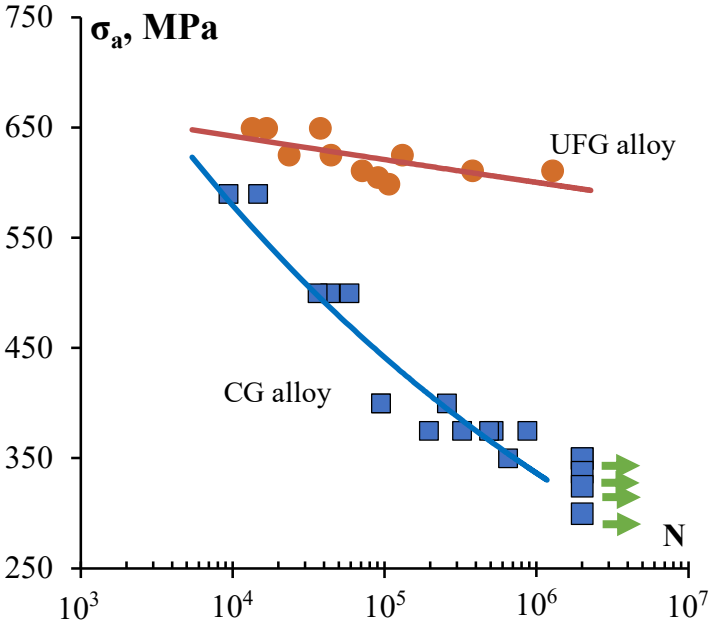

**Fig. 5** Results of corrosion-fatigue tests of coarse-grained and UFG Ti-5Al-2V alloy

Fig. 6 shows the results of SEM analysis of the fracture surfaces of the coarse-grained alloy samples after corrosion-fatigue tests. The fracture structure in Fig. 6 is typical for fractures of samples after fatigue tests (see [59]).

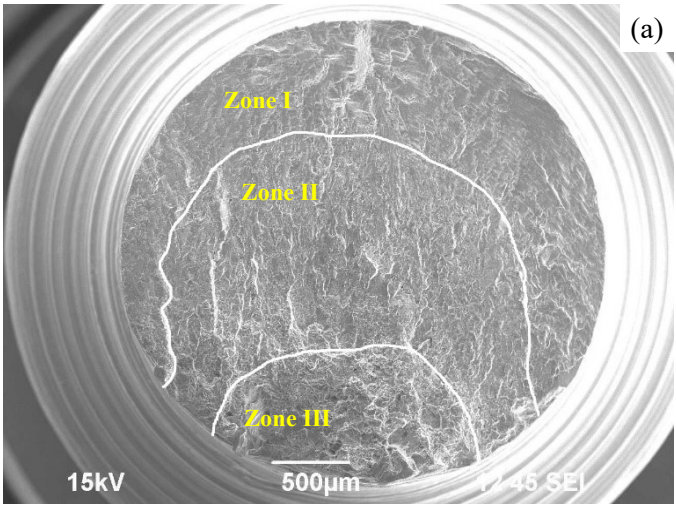

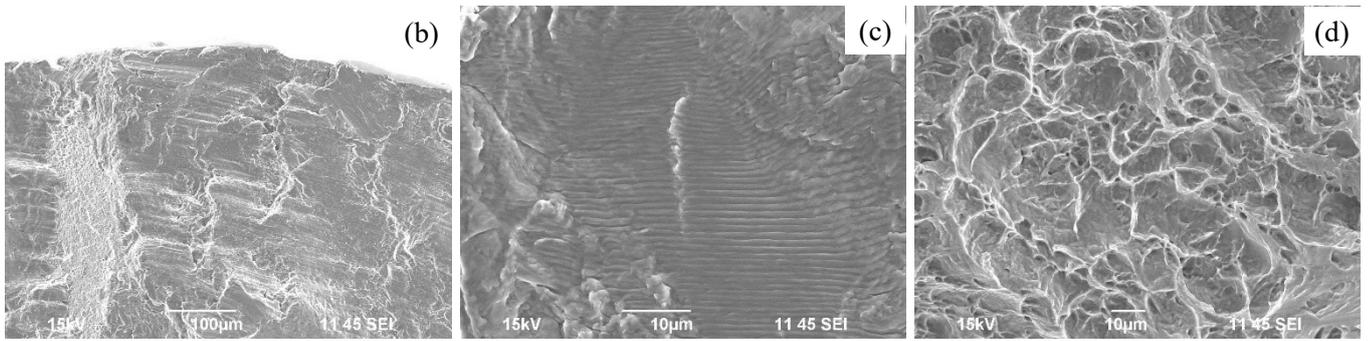

**Fig. 6** Fractographic analysis of the fracture surfaces of coarse-grained Ti-5Al-2V alloy samples: (a) general view of the fracture; (b) microcrack initiation area (zone 1 in Fig. 6a); (c) crack growth zone (zone 2 in Fig. 4a); (d) break region (zone 3 in Fig. 6a). Stress amplitude 500 MPa. SEM

Several characteristic zones can be distinguished on the fracture surfaces of the coarse-grained samples (Fig. 6a): the microcrack initiation zone (Zone #1), the crack growth zone (Zone #2), and the break zone (Zone #3). Depending on the stress amplitude magnitude, one or several microcrack initiation zones can be observed. At high stress amplitude magnitudes, typically only one crack initiation zone is observed. Study of Zone #2 indicates that the microcrack initiation occurs via the intragranular mechanism. The area of crack growth (Zone #2) is the largest one; the fracture surface at the stage of stable crack growth exhibits a mixed brittle-ductile nature. In Zone #2, the areas of brittle fracture (facet inside the grain, Fig. 6b) and the ones of ductile fracture (fatigue beachmarks, with an average spacing of ~ 1 μm, Fig. 6c) can be observed. In the break zone of the sample (Zone #3), numerous pits and dimples indicative of ductile fracture are observed (Fig. 6d).

The fatigue fractures of the UFG Ti-5Al-2V alloy samples have a more complex nature – Zone #2 is divided into a stable microcrack growth zone (Zone #2a) and an accelerated crack growth zone (Zone #2b) (Fig. 7). The fracture of the UFG alloy at the stage of stable crack growth exhibits a mixed brittle-ductile nature. Due to the small grain size, no intragranular facets are observed in the stable crack growth zone. In the Zone #2a, fatigue beachmarks are often observed (Fig. 7c). The distance between the fatigue beachmarks in the UFG samples is much smaller than in coarse-grained samples. The obtained result indirectly indicates that the growth rate of fatigue cracks in UFG alloys turns out to be much lower than in coarse-grained alloys. In the accelerated crack growth zone, secondary cracks are often observed (Fig. 7d). According to [59], the presence of the secondary cracks indicates ductile fracture in the accelerated crack growth zone. The area of Zone #3 in the UFG samples is much larger than in the coarse-grained samples. Numerous pits resulting from the merging of micro-pores can be observed in Zone #3, indicating the ductile nature of the fracture of the UFG material.

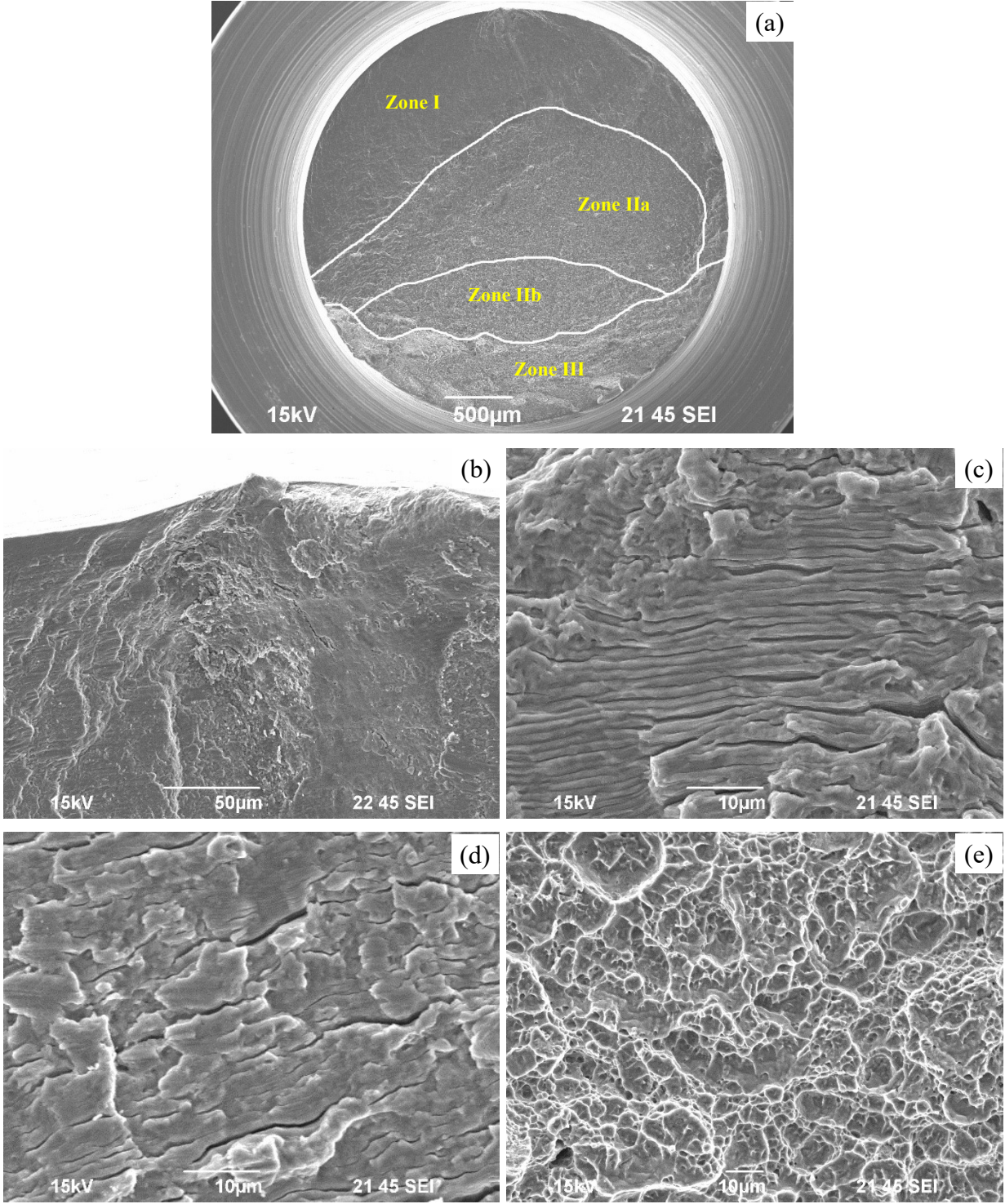

**Fig. 7** Fractographic analysis of the fracture surface of UFG Ti-5Al-2V alloy samples: (a) general view of the fracture; (b) microcrack initiation area (Zone #1 in Fig. 7a); (c) stable crack growth zone (Zone #2a in Fig. 7a); (d) accelerated crack growth zone (Zone #2b in Fig. 7a); (e) break region (Zone #3 in Fig. 7a). Stress amplitude 620 MPa. SEM

*3.2.2 Effect of annealing temperature*

Fig. 8 shows the $\sigma_a(N)$ dependencies for the UFG Ti-5Al-2V alloy samples subjected to 30-minute annealing at various temperatures. For ease of the analysis, the results of fatigue tests of the unannealed UFG Ti-5Al-2V alloy in the as-swaged state are also presented in Fig. 8. One can see from Fig. 8 and Table 1 the slopes of the $\sigma_a(N)$ curves to depend on the annealing temperature non-monotonously – the slope increased with increasing annealing temperature up to 600°C and decreased again after annealing at 700°C. At the same time, an increase in the annealing temperature resulted in a decrease in the fatigue limit and a reduction in the fatigue life of the alloy. The fatigue characteristics of the UFG Ti-5Al-2V alloy after annealing at 600 and 700°C are close to the ones of the coarse-grained alloy.

So far, the alloy with a partially recrystallized microstructure manifested an anomalous increase in the slope of the $\sigma_a(N)$ curve. In the UFG material with a fully recrystallized microstructure, the slope of the $\sigma_a(N)$ curve decreased again (Fig. 8). It is noteworthy that the non-monotonous nature of the change in the slope is quite an unexpected result. Traditionally, annealing is assumed to lead to a monotonous decrease in the parameter *q* in the Basquin equation. It should be noted also that the reduction in fatigue strength characteristics of the UFG titanium alloy Ti-5Al-2V begins at temperatures (500-550°C) lower than the grain growth start temperature (see Subsection 3.1).

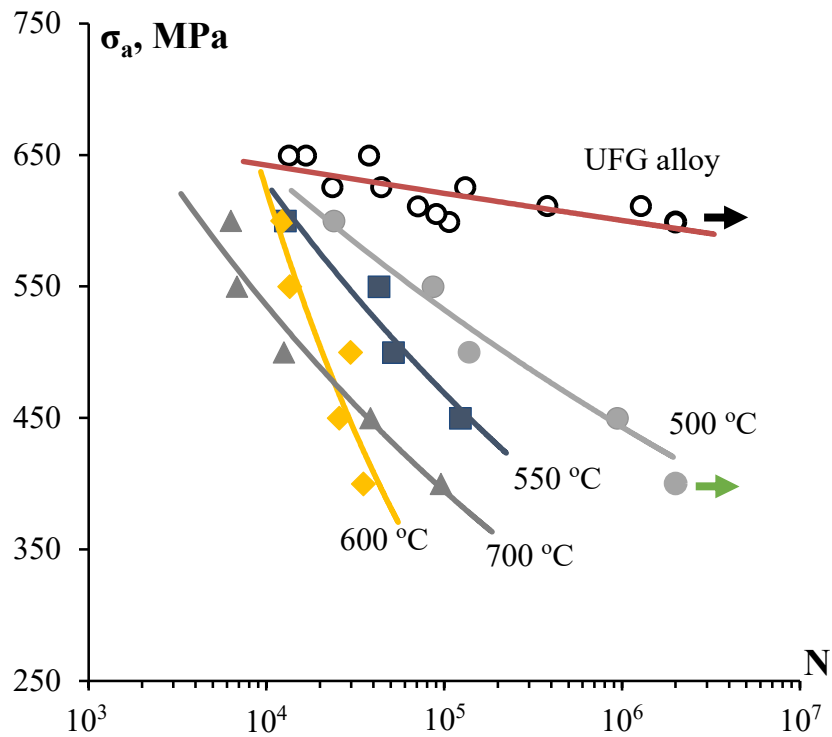

**Fig. 8** Results of corrosion-fatigue studies of annealed UFG alloys

Fig. 9 shows the results of fractographic analysis of fractures of annealed UFG samples after corrosion and fatigue tests. The analysis of the results of fractographic studies shows that on the surface of the fractures of annealed samples, an area of microcrack nucleation can be identified (Zone I in Fig. 9a, b), the crack growth zone (Zone II in Fig. 9a, b) and the break zone (Zone II in Fig. 9a, b). It is impossible to distinguish a sub-zone of stable crack growth and a sub-zone of accelerated crack growth in annealed samples. A ductility type of fracture is observed in the crack growth zone - secondary cracks are rarely found in Zone II and fatigue beachmarks are completely absent. The distance between the fatigue beachmarks in the UFG alloys annealed at 500-600 °C is very small - at the same resolution, the distance between the beachmarks is practically not visible even at high magnification.

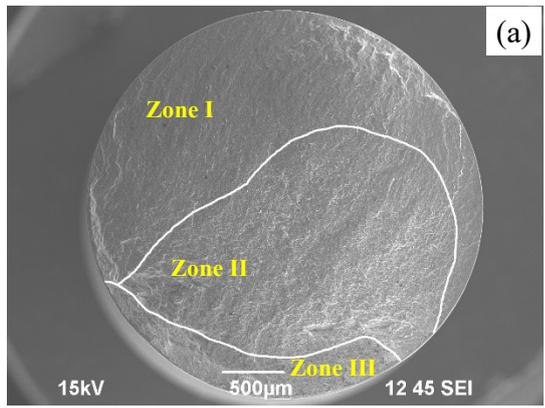

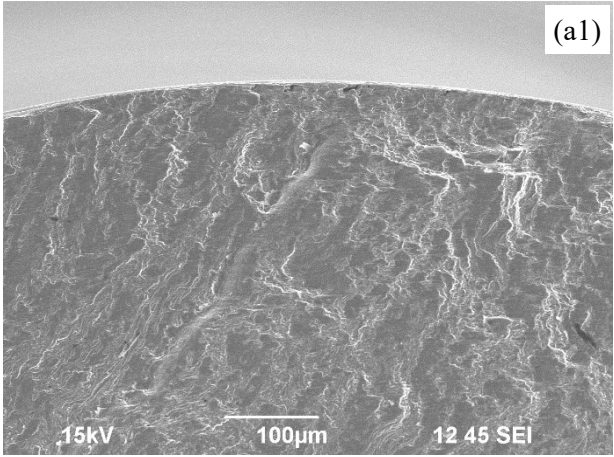 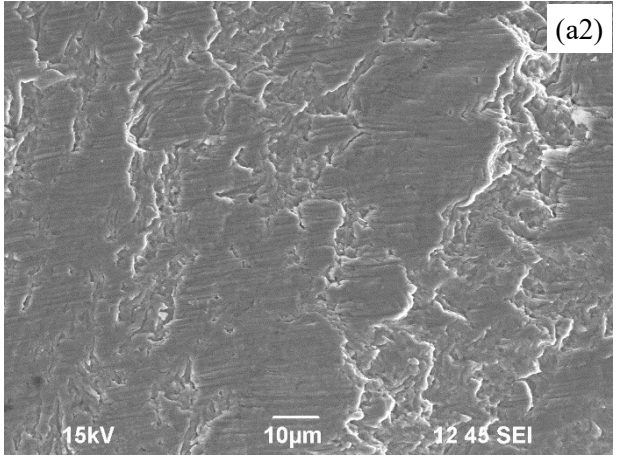

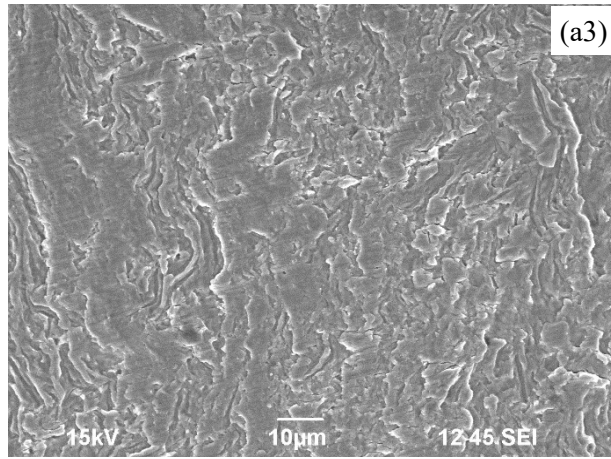 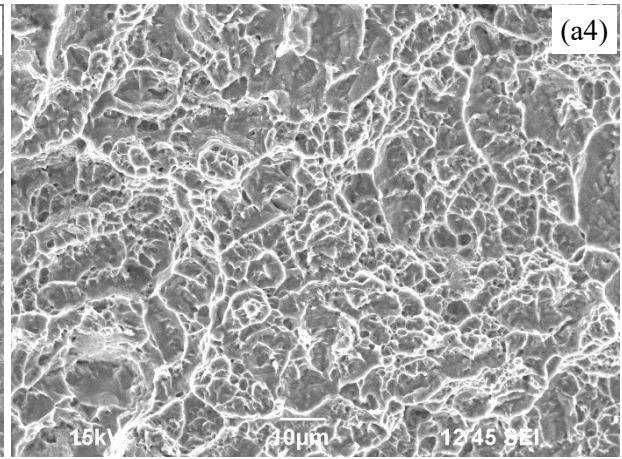

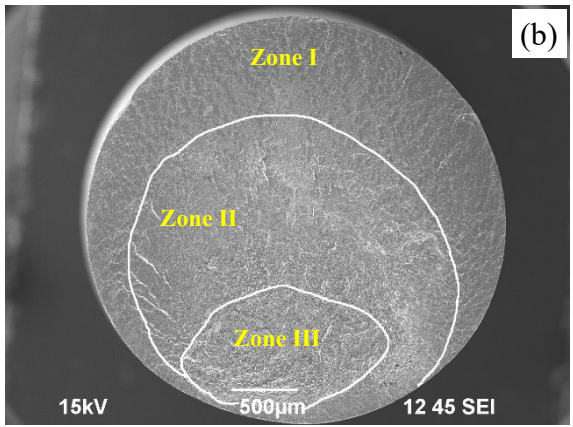

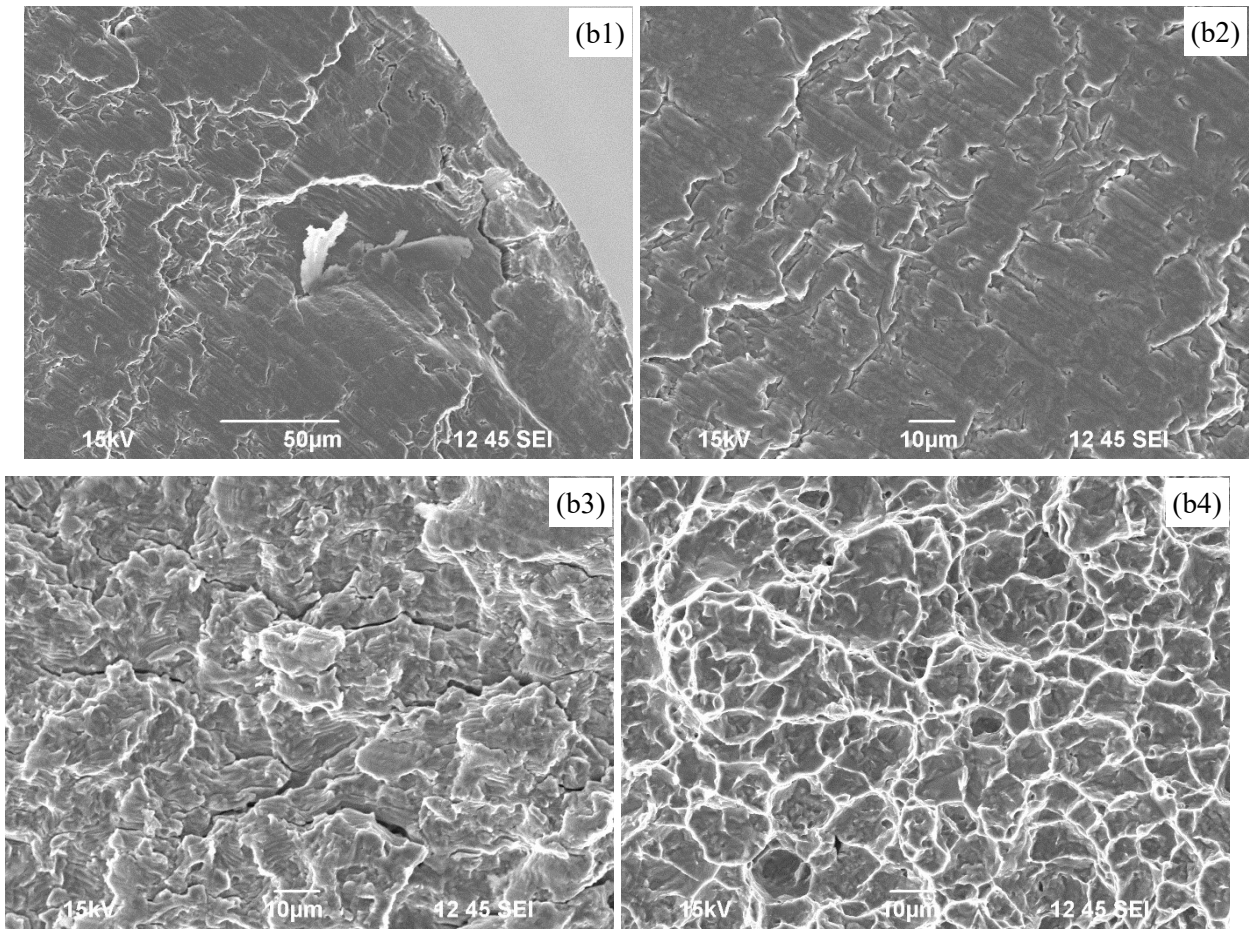

**Fig. 9** Fractographic analysis of fractures of UFG alloy samples after annealing at temperatures of 500 °C ($\sigma_a$ = 450 MPa) (a, a1, a2, a3, a4) and 700 °C ($\sigma_a$ = 400 MPa) (b, b1, b2, b3, b4): (a, b) general view of the fracture; (a1, b1) microcrack initiation area (Zone #1), (a2, a3, b2, b3) crack growth area (Zone #2); (a4, b4) break zone (Zone #4). SEM

*3.2.3 Effect of annealing time at 250 °C*

From Fig. 10, one can see the prolonged annealing of the UFG alloy Ti-5Al-2V at 250 °C not to affect its corrosion-fatigue strength characteristics significantly. The UFG alloy Ti-5Al-2V after prolonged annealing at 250 °C demonstrated high corrosion-fatigue strength indicating the high thermal stability of the UFG alloy microstructure. The parameters of the Basquin equation do not depend on the annealing time within the measurement uncertainty and are: $A$ = 790-825 MPa, $q$ = 0.02-0.03. The fracture characteristics of the samples annealed at 250 °C are identical to the ones of the unannealed UFG samples.

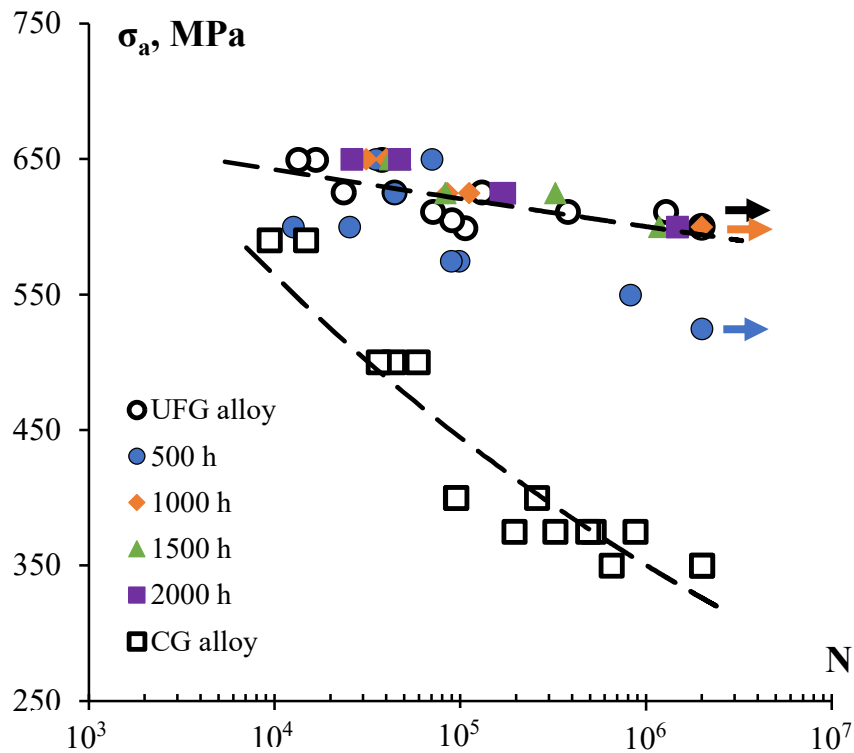

**Fig. 10** Results of investigations of the effect of annealing time at 250 °C on the fatigue strength of the UFG titanium alloy Ti-5Al-2V

3.3 Hot salt corrosion tests

*3.3.1 Tests of unannealed alloys*

A dense salt layer formed on the surfaces of the coarse-grained samples (Fig. 11a) showing good adhesive strength with the titanium alloy surfaces. The composition analysis of the surface layers using EDS revealed the corrosion products to consist of (in wt.%) 26.5% O, 23.8% Na, 1.4% Al, 0.2% S, 36.0% Cl, 11.7% Ti, and 0.4% V. This result indicates the surface layers of the titanium alloy samples after HSC tests to consist of NaCl salt, which includes sulfur microadditives as well as complex oxides of titanium, aluminum, vanadium, and intermetallic compounds based on them. After rinsing the sample in a stream of hot water, the salt crystals and easily soluble corrosion products dissolved leaving a dark-colored corrosion products on the sample surfaces - titanium, aluminum, and vanadium oxides as well as Ti-Al-V intermetallics (Fig. 11b).

The undissolved corrosion products were removed by chemical treatment according to GOST R 9.907-2004. Large corrosion pits approximately 1 mm in size were observed on the surfaces of the cleaned coarse-grained samples (Fig. 11c). The corrosion pits up to 0.5 mm deep were distributed evenly over the entire surface of the coarse-grained alloy. The average reduction in the cross-sectional area of the coarse-grained alloy sample after HSC testing did not exceed 4-5%.

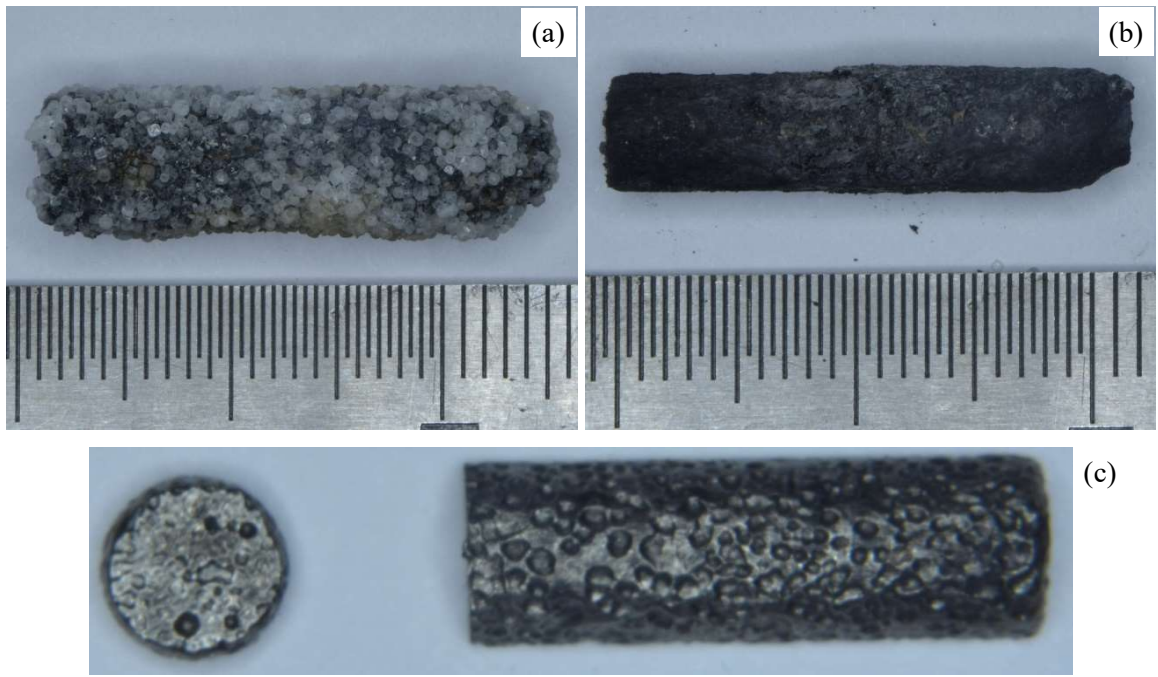

**Fig. 11** Character of the corrosion damage of the coarse-grained alloy Ti-5Al-2V specimen after HSC testing: (a) general view of the specimen after HSC testing; (b) general view of the specimen after rinsing in a hot water stream; (c) specimen surface after chemical removal of the corrosion products

The analysis of the XRD curves presented in Fig. 12a shows the composition of the products of HSC of the coarse-grained alloy Ti-5Al-2V to include the phases of the NaCl structure type (PDF #00-005-0628), titanium oxides (rutile $TiO_2$ (PDF #00-021-1276), which is the main phase, mon-TiO (PDF #01-072-0020), $Ti_2O_3$ (PDF #00-010-0063)), and $Ti_3Al$ (PDF #00-052-0859). In addition, the XRD curves manifested the peaks from α-Ti, the lattice parameters of which were increased as compared to the reference data (PDF #00-044-1294, ICSD #44390). It should be noted that in the XRD experiments it is not possible to distinguish between the phases $Ti_3Al$ (PDF #00-052-0859) and $Ti_{2.76}Al_{1.24}O_{0.35}$ (PDF #01-074-9927). So one can assume the corrosion products of the titanium alloy to contain both the $Ti_3Al$ ($a$ = 5.793 Å, $c$ = 4.649 Å) phase and $Ti_{2.76}Al_{1.24}O_{0.35}$ ($a$ = 5.787 Å, $c$ = 4.678 Å) one from the same space group P6/mmc.

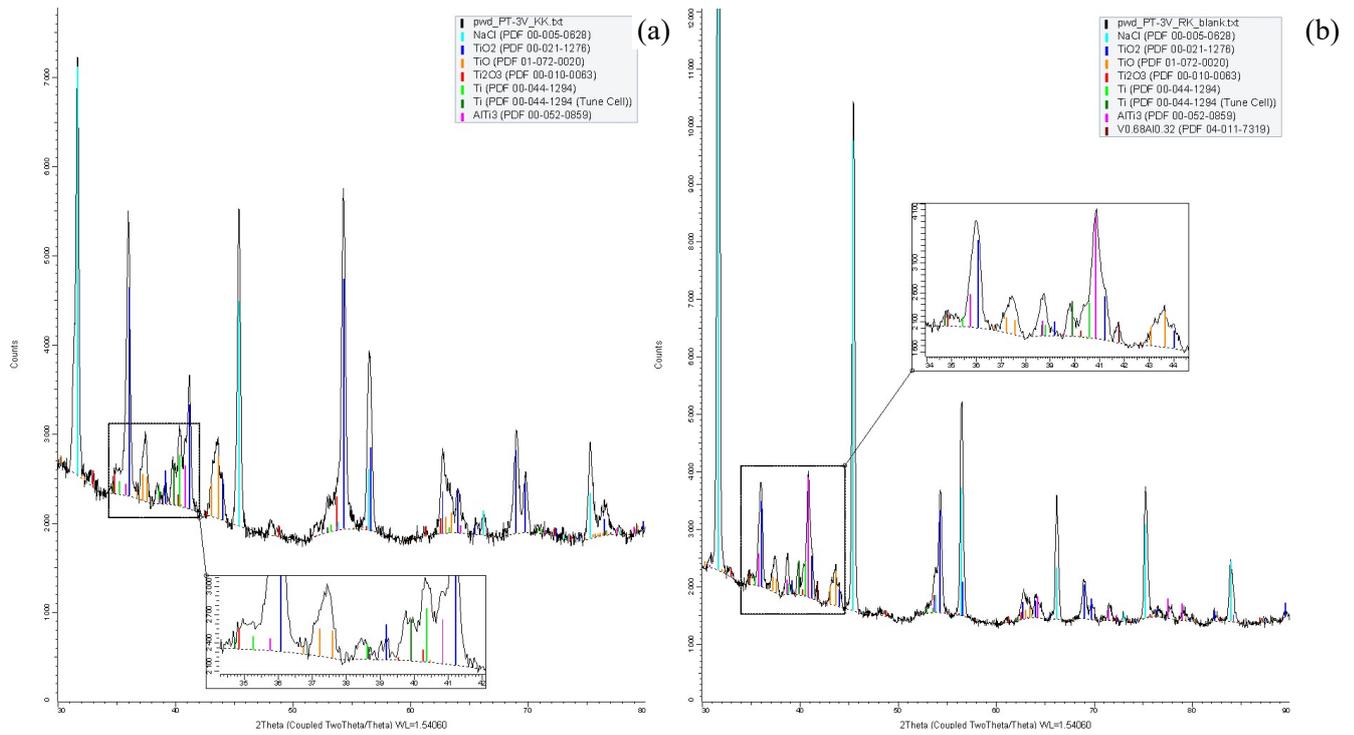

**Fig. 12** Results of XRD phase analysis of products of HSC of coarse-grained (a) and UFG alloy Ti-5Al-2V (b)

In the initial state, the coarse-grained alloy had an increased susceptibility to HSC. The micro-damage of the surfaces of the coarse-grained alloy goes via the intergranular corrosion (IGC) mechanism (Fig. 13). During the test, quite intensive destruction of the sample surfaces and destruction of the grain groups took place. This area is marked with a yellow dashed line in Fig. 13b. Subsequently, these areas are transformed into large corrosion pits (Fig. 13c, g). The maximum depth of IGC defects varied from 16 to 72 μm; the average depth of IGC defects was close to ~ 50 μm.

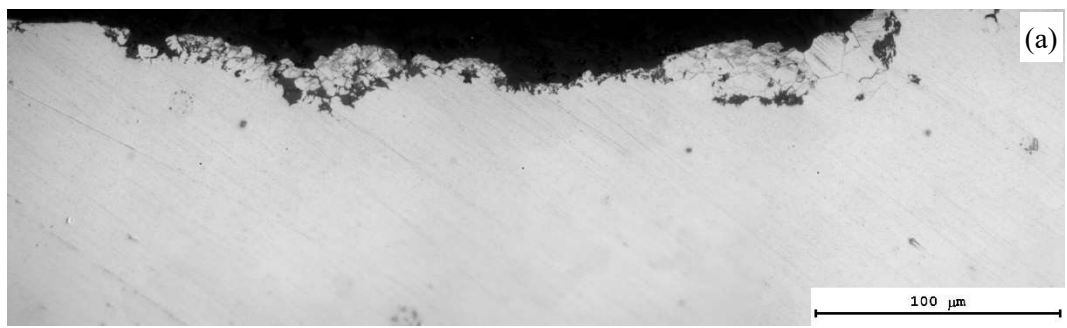

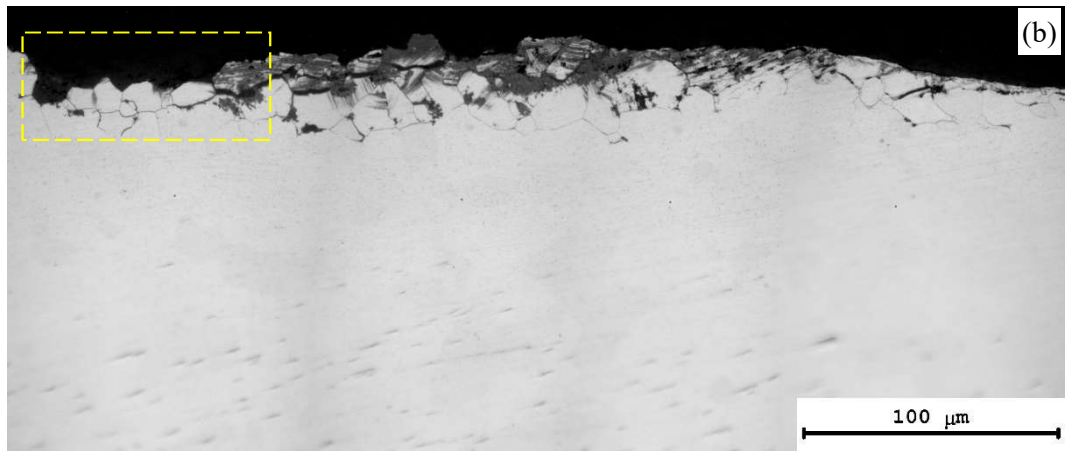

**Fig. 13** Character of the corrosion damage to the surface of PT-3V coarse-grained alloy specimens after HSC. Metallographic optical microscopy

Fig. 14 presents the results of metallographic optical microscopy examination of the nature of surface damage to the UFG Ti-5Al-2V alloy specimens after HSC testing. Surface damage to the UFG alloy goes via the IGC mechanism. The corrosion was quite uniform over the entire metal surface. The depth of the IGC defects reaches 100-120 µm. As the depth of the IGC defects increases, these ones transform into corrosion pits with a depth of 50-70 µm. The areas of transformation of the IGC defects into the corrosion pits are marked with yellow dashed lines in Fig. 14a, b. Thus, the HSC process of the alloy Ti-5Al-2V has a two-stage character: first, IGC occurs, and then the IGC defects transform into the corrosion pits. Afterwards, at further increasing of the test time, the IGC starts on again. As a result of this alternation of corrosion mechanisms, the cross-section area of the specimen decreases relatively uniformly as shown in Fig. 14c.

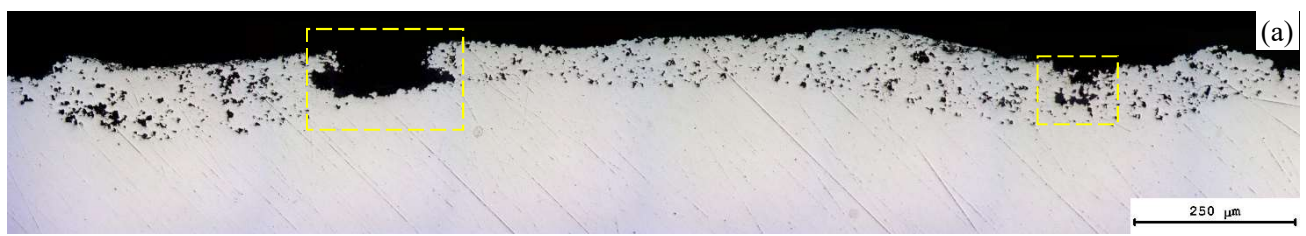

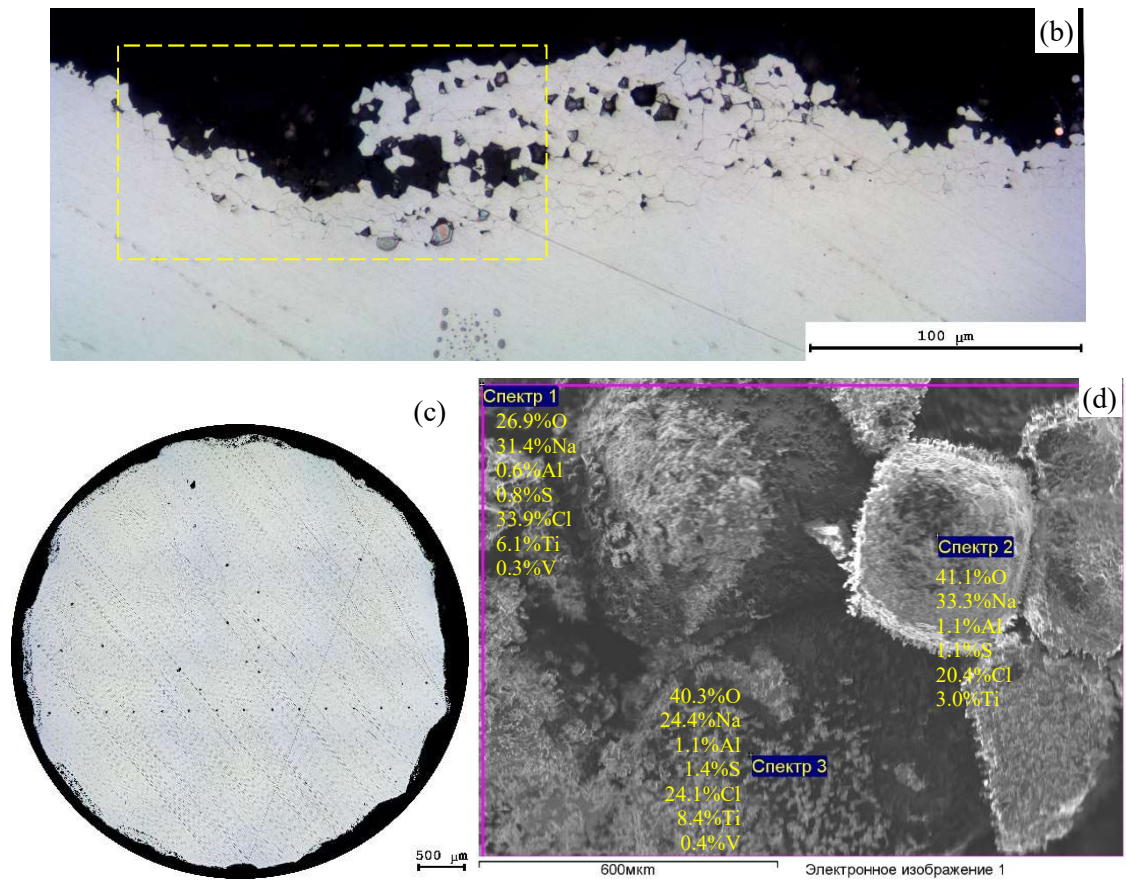

**Fig. 14** Results of the HSC tests of the UFG Ti-5Al-2V alloy specimens: (a) general view of the specimen surface with IGC defects; (b) area of transformation of the IGC defect into a corrosion pit; (c) cross-section of the specimen after HSC testing; (d) EDS analysis of the composition of the corrosion products

The composition of the HSC products on the surfaces of the UFG specimens (Fig. 14d) was similar to the ones for the coarse-grained alloy. The results of XRD phase analysis (Fig. 12b) show the phase composition of the corrosion products of the UFG alloy to be similar to that of the coarse-grained alloy. The XRD peak at 2θ ~ 41.7° is visible in the XRD curve of the corrosion products of the UFG alloy. This XRD peak may be ascribed to the phase $V_{0.68}Al_{0.32}$ (Im-3m, PDF #04-011-7319), which is a product of corrosion damage of grain boundaries enriched with vanadium (see [35, 48]).

The characteristic distance between the IGC defects in the UFG alloy specimens was 5-10 μm (Fig. 13b, 14b). The microhardness of the UFG alloy after the HSC tests decreased slightly down to 2890 ± 180 MPa while the microhardness of the coarse-grained alloy remained unchanged and was close to the hardness of the alloy in the initial state (Table 1). These results indicate that the process of recrystallization takes place in the titanium alloy specimen during HSC testing. As an example, Fig.15a, b present the results of the microstructure investigation of the UFG titanium alloy specimen after HSC testing (250 °C, 250 h). One can see from Fig. 15a, b the Ti-5Al-2V alloy specimens after HSC testing to have a recrystallized microstructure with average grain size of ~10 μm. The heterogeneity of the macrostructure, formed during RS, remains – lighter areas of recrystallized metal are visible in the sample volume (Fig. 15a). It is known that as the degree of strain increases, the size of the recrystallized grain decreases. The presence of areas of recrystallized metal in the central part of the bar indicates less strain during RS.

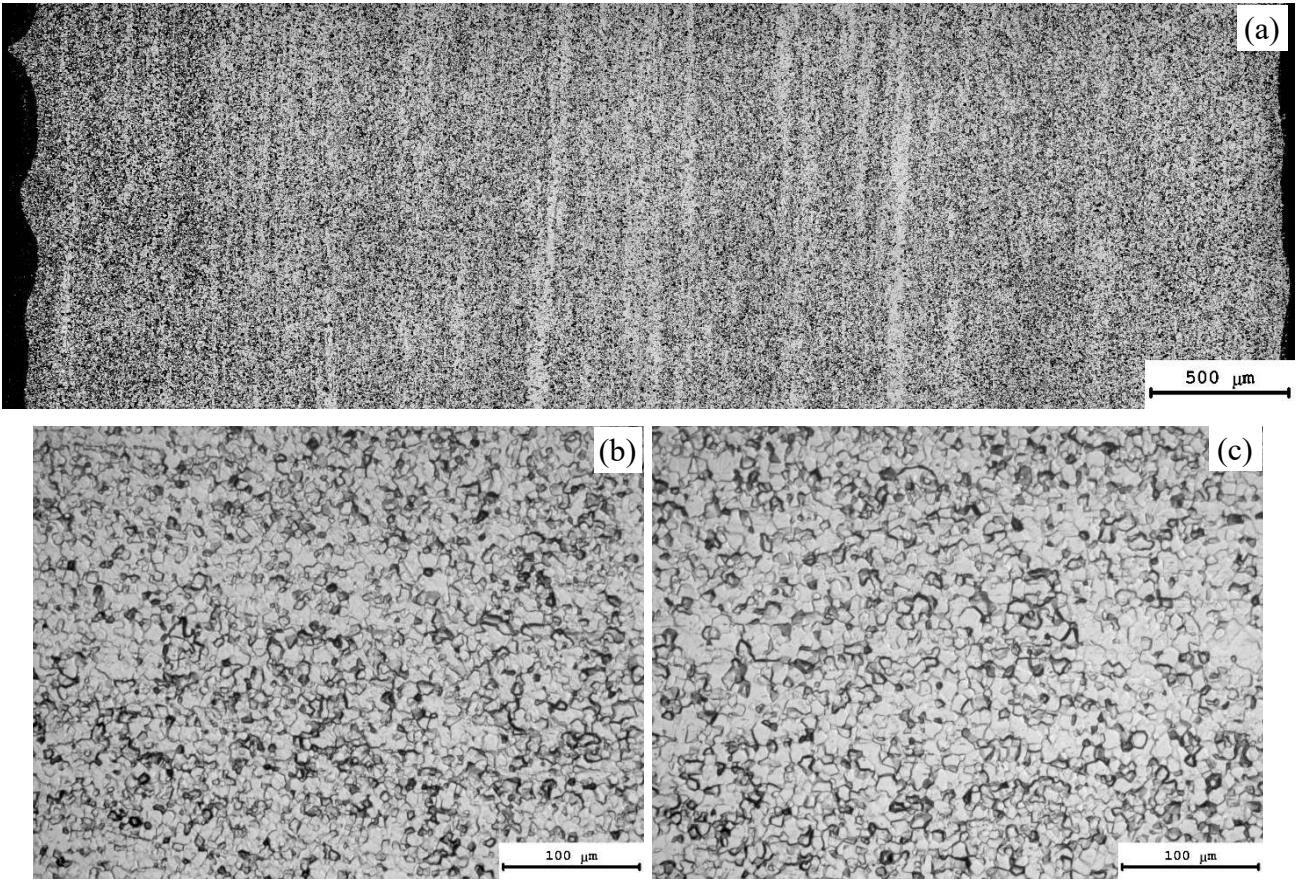

**Fig. 15** Microstructure of Ti-5Al-2V alloy specimens after HSC testing: (a, b) after RS, (c) after RS and annealing at 250 °C for 500 h

*3.3.2 Effect of the temperature of 30-minute annealing*

Let us analyze the nature of the corrosion damage to the specimens with partially recrystallized microstructure and fully recrystallized one.

The results of XRD studies indicate the annealing not to affect the phase composition of the corrosion products of the UFG titanium alloy Ti-5Al-2V.

Fig. 16 presents the results of the investigation of the nature of the corrosion damage to the UFG alloy specimens annealed at different temperatures. In the alloys with partially recrystallized microstructure, the local acceleration of IGC takes place (Fig. 16a) leading to the appearance of relatively small (50-100 μm) corrosion pits on the surfaces of the alloy specimens (Fig. 16b). The propagation rate of IGC exceeds the rate of formation and growth of the corrosion micro-pits. As a result, the IGC defects have a greater depth than the corrosion pits (Fig. 16b).

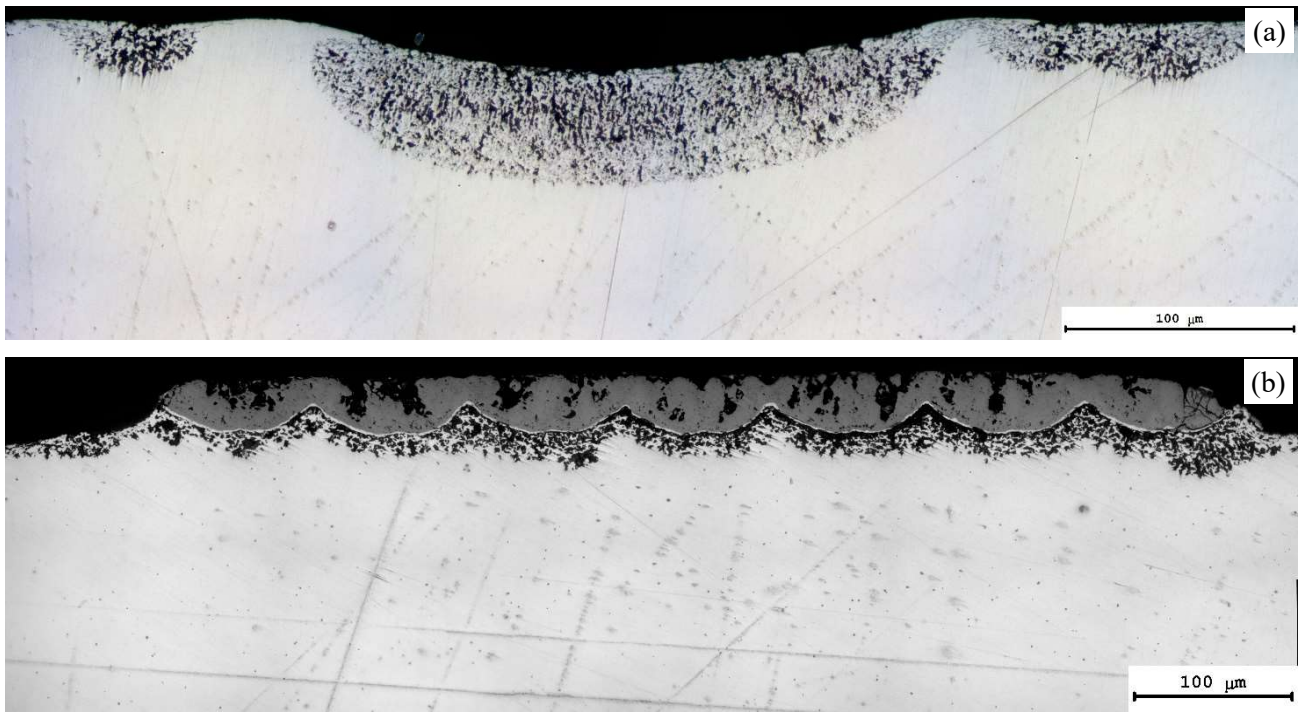

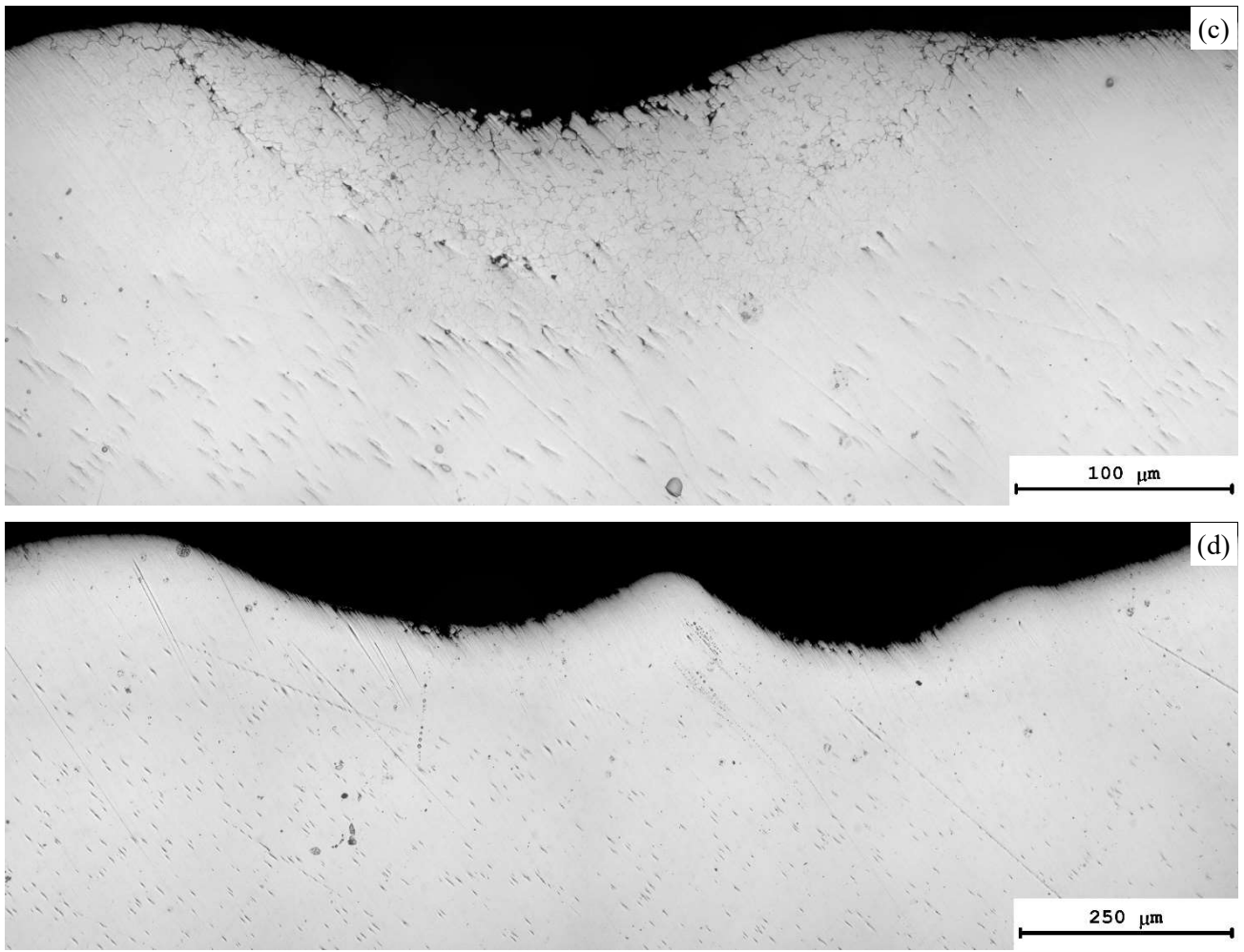

**Fig. 16** Character of the corrosion damage to the surface of UFG Ti-5Al-2V alloy specimens annealed at 550 °C (a), 600 °C (b), and 700 °C (c, d). Metallographic optical microscopy

On the surfaces of fully recrystallized specimens, the local IGC was present also (Fig. 16c) leading to the appearance of the corrosion pits up to 300-500 μm in size (Fig. 16d). The measurements of the corrosion defects' depths have shown an increase in the annealing temperature to result in a reduction in the IGC depth (Fig. 17). It was noted also that the alloys with fully recrystallized microstructure exhibit an increased overall corrosion rate. In the alloys with partially recrystallized microstructure, the reduction in the cross-sectional areas of the specimens after HSC testing was 6-8% while in the specimens annealed at 700 °C, it was 11-12%. The higher corrosion pit growth rate and overall corrosion rate lead to a decrease in the number of IGC defects on the surfaces of the specimens with fully recrystallized microstructure.

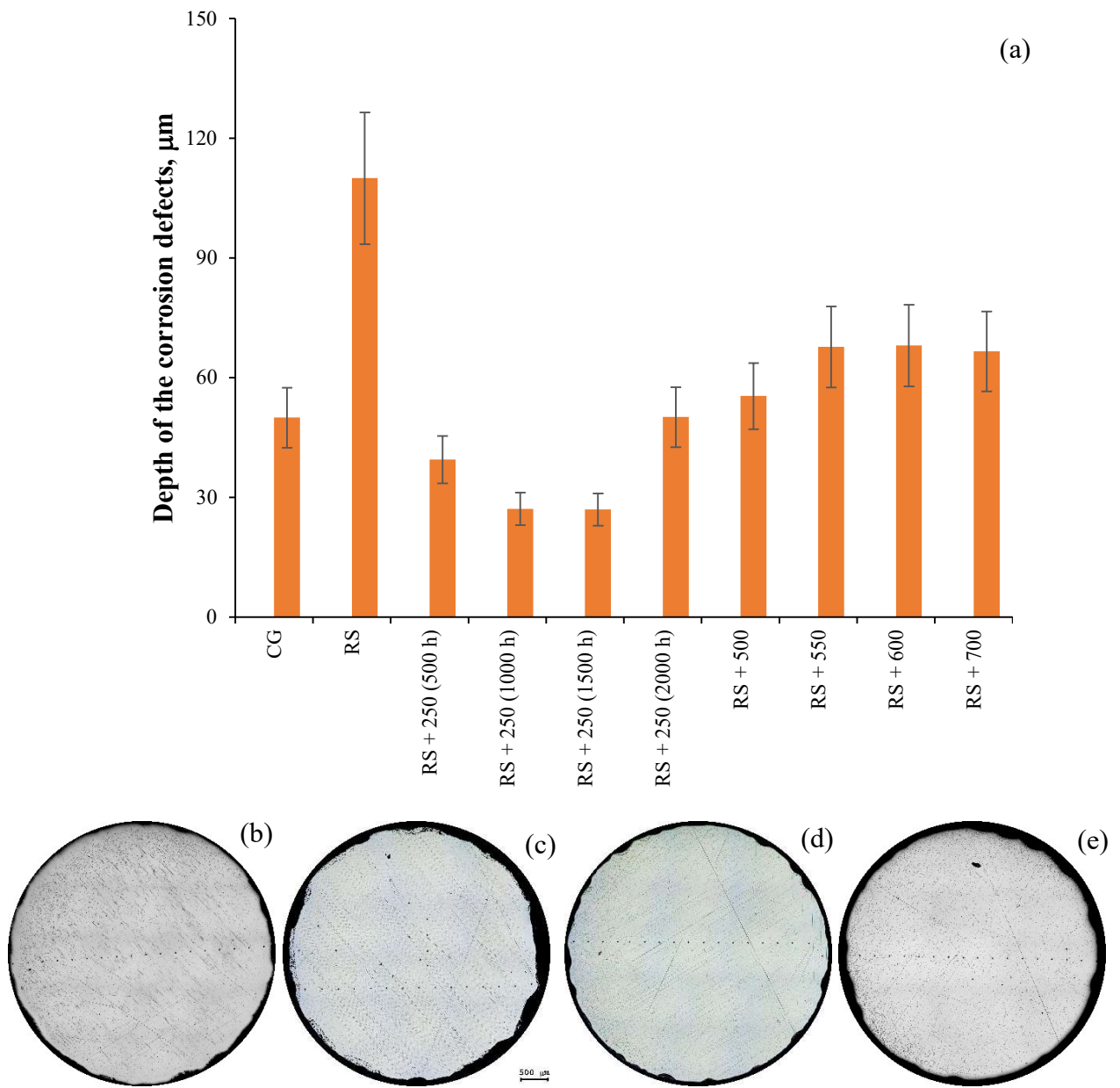

**Fig. 17** Depth of the defects in titanium specimens after HSC testing (a) and images of the cross-section of samples after testing on HSC (b-e): (b) coarse-grained sample, (c) un-annealed UFG sample, (d) UFG sample after annealing by, (e) UFG sample after annealing by 700 °C

*3.3.3 Effect of annealing time at 250 °C*

The results of HSC testing show the annealing time at 250 °C not to affect noticeably the mechanism and corrosion rate of the UFG titanium near-α alloy Ti-5Al-2V specimens. The phase composition of the HSC products of the UFG specimens annealed at 250 °C matches to the one of the unannealed UFG alloy specimens. As one can see from the comparison between Fig. 18a and Fig.

18b, the IGC develops during HSC testing, which then transforms into the submicron-scale corrosion pits. The average depth of the IGC defects' penetration does not depend on the annealing time at 250 °C within the measurement uncertainty (Fig. 18c, d). The average size of the fragment surrounded by the IGC defects was approximately 10 µm (Fig. 18c, d) that is close to the grain size in the alloy (Fig. 15c).

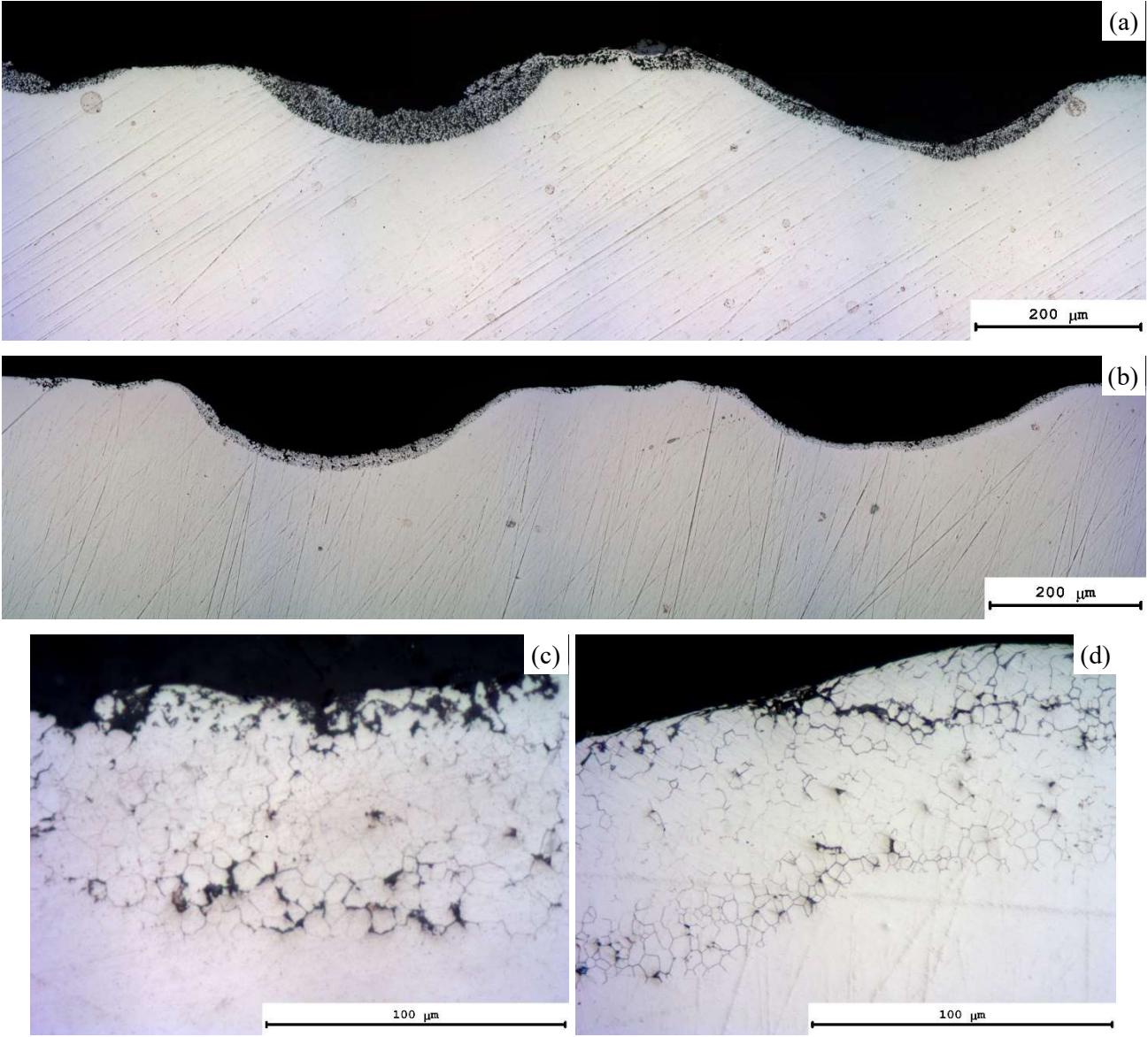

**Fig. 18** Character of the corrosion damage to the surfaces of UFG Ti-5Al-2V alloy specimens after HSC testing: (a, c) annealing at 250 °C, 500 h; (b, d) annealing at 250 °C, 2000 h

After HSC testing, the microhardness of all specimens ranged from 2540-2600 MPa to 2900-2940 MPa. The microhardness of the central layers of the specimens was approximately 300 MPa lower than the one of the surface layers.

**4. Discussion**

4.1 Microstructure and strength of the near-α titanium alloy

Earlier, two types of grain boundaries were shown in coarse-grained titanium alloy Ti-5Al-2V [35]: (i) conventional grain boundaries, where the concentrations of aluminum and vanadium are close to the ones in the grains' crystal lattice; and (ii) grain boundaries with an increased local vanadium concentration (up to 10 wt.%) containing β-phase particles often (Fig. 19). According to the results of EDS analysis, the β-phase particles in the Ti-5Al-2V alloy contain up to 16 wt.% V, as well as minor concentrations of aluminum and iron [35] (Fig. 19). The volume fraction of the second-type V-enriched grain boundaries was not large and was approximately 10%. The presence of two types of grain boundaries in the microstructure of the Ti-5Al-2V alloy leads to the appearance of two types of IGC defects during HSC testing – thin, shallow defects propagating along the grain boundaries of the first type, and thick, deep IGC defects associated with the grain boundaries of the second type.

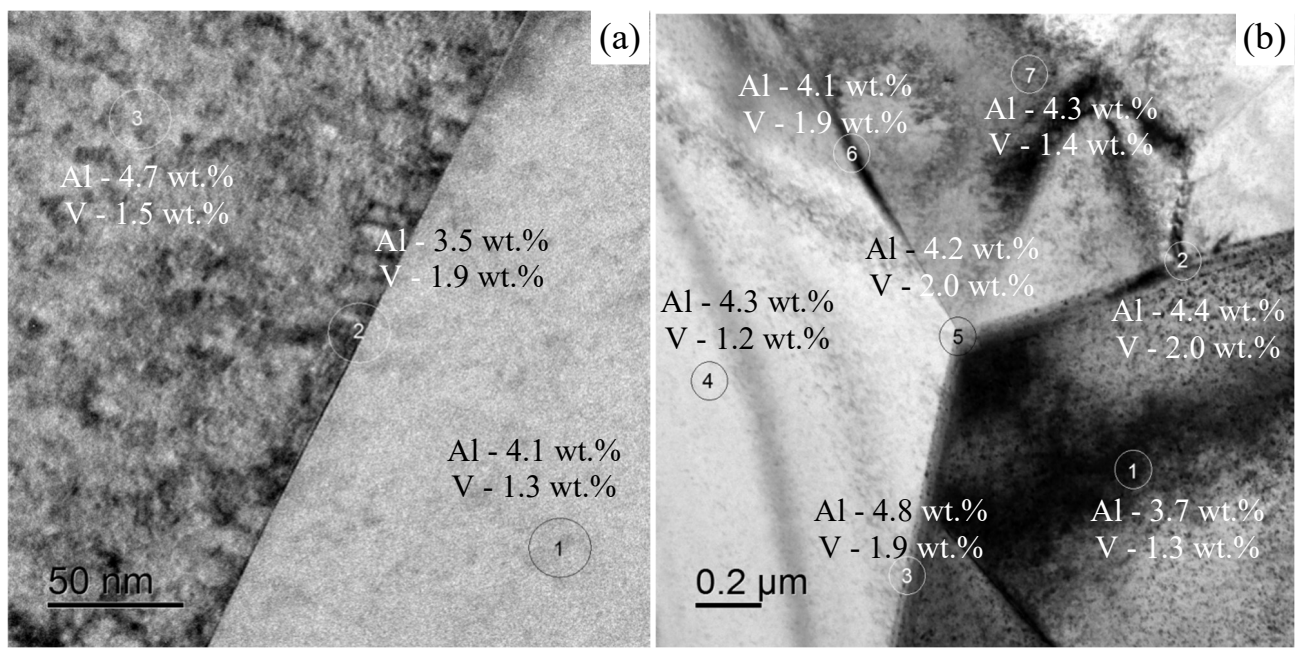

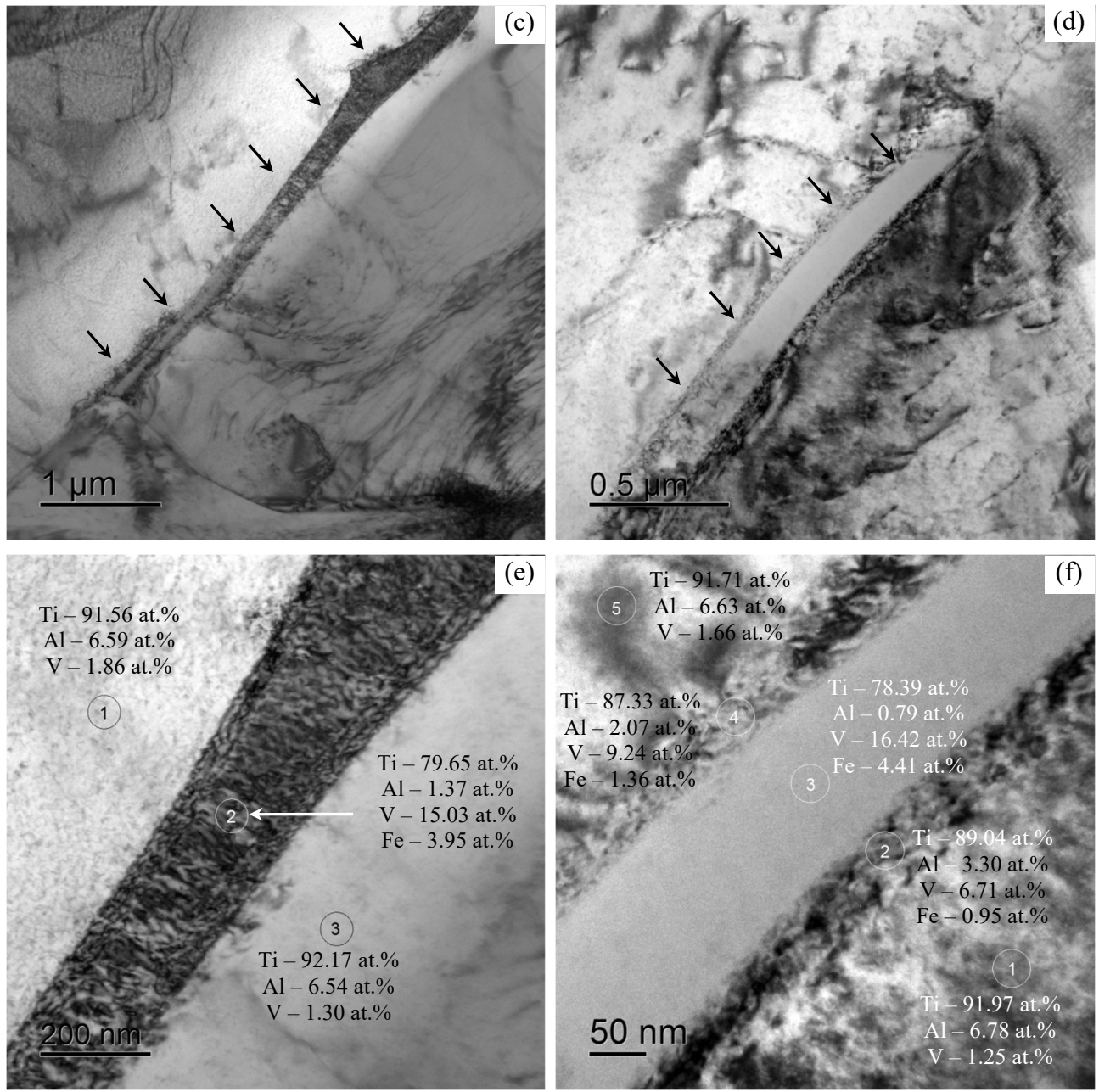

**Fig. 19**. Microstructure of the coarse-grained alloy Ti-5Al-2V: (a, b) conventional grain boundaries (type I) (c, d) precipitation of the β-phase particles (TEM) (β-phase particles are shown with arrows); (e, f) EDS results for the grain boundaries composition (type II) [35]

The strain during cold RS of titanium alloys was approximately 70%. This is a small amount of strain compared to severe plastic deformation (SPD) methods such as Equal Channel Angular Pressing / Extrusion (ECAP / ECAE) or High Pressure Torsion (HPT). In the RS process, a greater amount of plastic deformation energy is assumed to be expended on increasing the lattice dislocations density and the formation of submicron recrystallization nuclei. In this case, the microstructure of the Ti-5Al-2V alloy after RS can be described as follows. Heavily deformed α-Ti grains containing a high density of lattice dislocations ($\rho_v$), surrounded by old (primary) grain boundaries of the coarse-grained alloy. Since the deformation occurs at room temperature, there is no change in the local vanadium concentration at the grain boundaries. β-phase particles located at the boundaries of α'-grains are fragment during plastic deformation realized by RS.

The contribution of the lattice dislocations into the macroelasticity stress (lattice friction stress) of a UFG metal can be estimated using the equation [60]:

$$\Delta\sigma_\perp = \alpha G b \sqrt{\rho_v}, \qquad (1)$$

where $\alpha = 0.1\text{-}0.2$ is a numerical coefficient, $G = 43.6$ GPa is the shear modulus of titanium, $b = 2.95 \cdot 10^{-10}$ m is the Burgers' vector of α-Ti. Assuming the macroelasticity stress to increase ~1.1 times (from 370 to 415 MPa, see Table 1) during RS, one can derive a slight increase in the lattice dislocations density (~1.17 times) to be necessary to achieve the increase in the strength observed. This is a typical scale of increase in the dislocation density for cold plastic deformation [38, 60]. Therefore, the main contribution to the increase in macroelasticity stress (lattice friction stress) of the titanium alloy Ti-5Al-2V during RS is due to the increase in the lattice dislocations' density.

The effect of grain boundary hardening to the yield strength of the Ti-5Al-2V alloy is not very large, since simultaneously with a decrease in grain size, the Hall-Patch coefficient decreases from 1.9-3.0 to 0.45-0.5 MPa·m$^{1/2}$. The decrease in the K coefficient is due to the grinding of the β-phase particles located along the grain boundaries of the coarse-grained alloy (see above).

Prior to analyzing the results, it should be noted that titanium is characterized abnormally low grain boundary diffusion activation energy [61]. As a result, the diffusion processes in titanium is controlled by grain boundaries actively and occur at temperatures $T < 0.5\ T_m$, where $T_m = 1943$ K is the melting point of titanium. As the analysis of the results shows, after prolonged 2000-hour holding at 250 °C under conditions of hot salt corrosion, grain growth was observed (Fig. 15).

After heating up to 550-600°C, the recrystallization process and the growth of the recrystallization nuclei begins. The low grain growth rate during high temperature annealing is due to the following two factors. First, upon heating the Ti-5Al-2V titanium alloy, titanium carbide TiC nanoparticles are precipitating [35]. This leads to an increase in the yield strength of the material at 500-550°C. Second, the 'primary' grain boundaries of the titanium alloy contain elevated concentrations of vanadium [35]. This prevents the formation of the microstructure through annealing when the recrystallized grain sizes exceed the original ones in the coarse-grained alloy.

Attention should be drawn to the abnormally reduced yield strength of the Ti-5Al-2V alloy annealed at 700°C (Table 1). Since the average size of the recrystallized grains was smaller than the grain sizes in the coarse-grained alloy and the contribution of the lattice dislocations to the yield strength of titanium is small (as mentioned above), one could expect the yield strength of the annealed UFG alloy to be greater than that of the coarse-grained alloy. However, the analysis of the results obtained shows the yield strength of the UFG alloy annealed at 700°C (~665 MPa) to be smaller than the one of the coarse-grained alloy (~735 MPa). In our opinion, the abnormal reduction in the yield strength of the recrystallized metal is associated with the significant fragmentation of the β-phase particles located along the boundaries of α'-grains in the titanium alloy. Large β-phase particles are known to enhance the strength of titanium alloys significantly [62]. The absence of β-phase particles along the grain boundaries does not allow for increased values of the Hall-Petch coefficient. The Hall-Petch coefficient $K = (\sigma_y - \sigma_0)\sqrt{d}$ after annealing of the UFG alloy at 700 °C is 0.7-1 MPa·m$^{1/2}$, in coarse-grained alloy - $K$ = 1.9-3.0 MPa·m$^{1/2}$.

### 4.2 Corrosion-fatigue testing

The comparative analysis of the test results for the coarse-grained and UFG alloys shows the UFG alloy to have a higher fatigue limit and increased cyclic durability (Fig. 4). As it is well known, the severe plastic deformation and strain hardening leads to an increase in the fatigue properties of the metal alloys. In this regard, one can consider the result obtained to be well known.

The effect of annealing temperature on the coefficient $A$ in the Basquin equation is quite unusual. As one can see from Fig. 8 and Table 1, the increase in the slope of the $\sigma_a(N)$ curve was observed in the UFG alloy with partially recrystallized microstructure, and after annealing at 700°C, the slope of the $\sigma_a(N)$ curve decreases again. It should be noted also that the value of coefficient $A$ in the UFG alloy was smaller than in the coarse-grained alloy (Table 1).

Note in advance also that a neutral corrosive environment does not affect the fatigue characteristics of smooth specimens of the titanium near-α alloy Ti-5Al-2V significantly [63]. However, the neutral corrosive environment affects the fatigue characteristics of notched specimens or titanium specimens with IGC defects essentially [30, 31, 63, 64]. In the present study, smooth cylindrical specimens without notches were tested. Therefore, the effect of the corrosive environment can be neglected in the first approximation. In accordance with the model of plastic initiation and growth of fatigue microcracks, the value of coefficient $A$ corresponds to the plastic deformation activation energy $\Delta F$ [37]. Thus, one can conclude that the reduction in the plastic deformation activation energy takes place in the UFG near-α alloy Ti-5Al-2V after RS. It should be noted that processing the α-alloy Ti-2.5Al-2.5Zr by RS led to an increase in coefficient $A$ and, consequently, the activation energy $\Delta F$. In our opinion, higher values of coefficient $A$ in the coarse-grained alloy are associated with the presence of large β-phase particles located along the boundaries of α′-grains in the titanium near-α alloy Ti-5Al-2V (Fig. 1b). The β-phase particles hinder the movement of lattice dislocations and impede the process of dislocation transfer from one grain to another. During cold severe plastic deformation, significant fragmentation of the β-phase particles took place, and the elongation of the titanium billet during RS led to an increase in the distance between fragmented β-phase particles. This reduces their ability to hinder the transfer of plastic deformation from one grain to another.

From Table 1, one can see the start of recrystallization and the start of grain boundary migration to lead to an increase in the value of coefficient $A$ (i.e. an increase in activation energy $\Delta F$). Earlier, the migrating grain boundaries in severely deformed metals were shown to accumulate dislocations that were located initially within the grain volume [60]. The intensity of the dislocation flow entering the migrating grain boundary is equal to $I_v = \varphi_1 \rho_v V_m$ where $\varphi_1$ is a numerical coefficient, $\rho_v$ is the dislocation density, and $V_m$ is the grain boundary migration velocity [60].

The defects captured by the grain boundaries lead to the appearance of the orientation mismatch dislocations (OMDs), which are characterized by the density $\rho_b \Delta b$ [60]. In the course of heating, OMDs can break up forming tangential components of the Burgers vector ($w_t$) [60]. The accumulating OMDs ($\rho_b \Delta b$) and tangential components of the Burgers vector ($w_t$) in the migrating grain boundaries lead to the formation of long-range internal stresses from the grain boundaries $\sigma_{gb} = \alpha_1 G \rho_b \Delta b + \alpha_2 G w_t$, where $\alpha_1$ and $\alpha_2$ are the numerical coefficients [60]. The long-range internal stress fields hinder the slip of lattice dislocations near the grain boundaries and, therefore, lead to an increase in the plastic deformation activation energy $\Delta F$. An increase in the annealing temperature leads to the formation of fully recrystallized microstructure with low lattice dislocation density $\rho_v$. This leads to a reduction in the flow of lattice dislocations entering grain boundaries ($I_v$), the long-range internal stress fields from the grain boundaries ($\sigma_{gb}$), and the plastic deformation activation energy ($\Delta F$), as it was observed in the experiment.

The second factor that can lead to an increase in the coefficient $A$ after annealing of the Ti-5Al-2V UFG alloy at 600 ºC may be the release of titanium carbide particles (see [35]). The precipitated nanoparticles prevent the slip of lattice dislocations and, thereby, contribute to an increase in the activation energy of plastic deformation $\Delta F$.

Now let us discuss the reasons for the decrease in the fatigue characteristics of the fully recrystallized Ti-5Al-2V alloy as compared to the characteristics of the coarse-grained alloy (Table 1, Fig. 8). As it was shown earlier, recrystallization annealing leads to the formation of an equilibrium microstructure, where the grain boundaries have a low density of excess defects. In addition, the grain boundaries of the recrystallized titanium alloy are free from large β-phase particles (Fig. 3), which are present in the grain boundaries of the coarse-grained alloy (Fig. 1b). This leads to rapid accumulation of dislocations at the grain boundaries during cyclic loading and, consequently, a reduction in the number of cycles to fatigue crack initiation.

### 4.3 Hot salt corrosion tests

From Figs. 13, 14, and 18, one can see that HSC in near-α titanium alloy Ti-5Al-2V specimens has an intergranular character. The key factors affecting the tendency of titanium alloys Ti-Al-V to hot salt intergranular corrosion (under the specified conditions of HSC testing) are: (i) the presence of the β-phase particles, which contain an increased concentration of alloying elements stabilizing the β-Ti phase at room temperature and (ii) an increased concentration of alloying elements at the grain boundaries, primarily vanadium [6, 35, 48].

The β-phase particles affect the susceptibility of the Ti-5Al-2V titanium alloy to IGC more significantly. Cold RS leads to the refinement of the β-phase particles but does not affect the chemical composition of the grain boundaries in titanium alloys significantly. Hence, one could expect the refinement of β-phase particles during RS to improve the corrosion resistance of the alloy. The analysis of the HSC test results shows the depth of IGC defects in the coarse-grained Ti-5Al-2V alloy to be approximately 50 μm, whereas in the UFG alloy after RS, it exceeded 100 μm.

In our opinion, the main origin of the increased susceptibility of severely deformed Ti-5Al-2V alloy to IGC is the particular conditions of the HSC tests. It should be noted that the HSC tests were conducted at elevated temperature (250 °C) that leads to the initiation of recrystallization and grain growth in the UFG Ti-5Al-2V alloy. It is also important to note that vanadium is a horophilic doping element in titanium and tends to segregate at the grain boundaries of α-Ti [35].

The formation of new recrystallization nuclei during RS and their growth during HSC tests leads to the emergence of novelty (secondary) grain boundaries in the recrystallized material containing an increased concentration of vanadium. These V-enriched grain boundaries serve as additional areas for intergranular corrosion development in the titanium alloy during HSC testing. This hypothesis is consistent with the results of microstructure studies of titanium alloys – as it is evident from the comparison of Fig. 3 and 15 with Fig. 14, 16, and 18, the average distance between the IGC lines corresponds to the grain size in the recrystallized Ti-5Al-2V alloy. Increasing the temperature or annealing time will lead to an increase in the average grain size and, consequently, a decrease in the area of vanadium-enriched grain boundaries. This results in reduced IGC depth with increasing temperature or annealing time of the deformed Ti-5Al-2V alloy (Fig. 17).

In conclusion, it is necessary to discuss the reasons for the increased tendency to pitting corrosion of samples of the Ti-5Al-2V UFG alloy obtained by the RS method. It can be seen from Fig. 11 and Fig. 18 that large corrosion pits are present on the surface of the samples after HSC tests. The reason for the formation of large corrosion pits is the heterogeneous macrostructure of the Ti-5Al-2V titanium alloy, which is formed during long-term tests on HSC. During the HSC test, a recrystallization process takes place and in the volume of the titanium rod, areas of recrystallized metal alternate with areas of fine-grained metal with an increased density of defects (Fig. 15a). As a result, macroscopic regions with different corrosion resistance are formed in the volume of the titanium pseudo-α alloy during the HSC test. The presence of such macroscopic areas will contribute to the accelerated formation of corrosion ulcers during HSC tests.

**Conclusions**

1. Rotary Swaging (RS) leads to an increase in the corrosion-fatigue strength of the Ti-5Al-2V titanium near-α alloy, resulting in enhanced fatigue limit and cyclic durability. The slopes of the fatigue curves $\sigma_a(N)$ (parameter *A* in the Basquin equation) for the deformed Ti-5Al-2V alloy is lower than the one for the coarse-grained alloy. Annealing leads to a non-monotonous change in parameter *A* – the alloy with partially recrystallized microstructure has higher values of the coefficient *A* that is due to the accumulation of defects at the migrating grain boundaries and the formation of long-range internal stress fields. The fully recrystallized alloy has low values of the coefficient *A* and low corrosion-fatigue strength. The absence of large β-phase particles at the grain boundaries of the recrystallized alloy does not allow achieving high fatigue limit and higher cyclic durability characteristics of the Ti-5Al-2V alloy.

2. Hot salt corrosion tests of the Ti-5Al-2V titanium near-α alloy were conducted in coarse-grained, fine-grained (after RS), partially recrystallized, and fully recrystallized states. Hot salt corrosion of the Ti-5Al-2V alloy has been shown to occur through an intergranular corrosion mechanism. XRD phase analysis results indicate vanadium to be present in the products of hot salt corrosion of deformed alloys, which is a horophilic alloying element in titanium and is concentrated along the grain boundaries predominantly. The emergence of new grain boundaries with increased vanadium concentration leads to a reduction in the corrosion resistance of the deformed Ti-5Al-2V alloy. Reducing the area of vanadium-enriched grain boundaries during annealing results in a reduction in the depth of intergranular hot salt corrosion defects.

3. The prolonged 2000-hour annealing at 250° C has been demonstrated not to result in any reduction in the corrosion-fatigue strength and corrosion resistance of the heavily deformed Ti-5Al-2V alloy. It was noted that during hot salt corrosion testing, recrystallization of the deformed alloy occurs resulting in a slight decrease in the microhardness.

**CRediT Authorship Contribution Statement:** V.N. Chuvil'deev - Formal analysis, Methodology, Writing - review and editing, Project administration, Funding acquisition, Supervision, Resources; A.A. Murashov & N.N. Berendeev – Investigation (Fatigue test), C.V. Likhnitskii – Investigation (Metallography, Microhardness), A.N. Sysoev & N.V. Melekhin – Investigation (Compression tests), K.A. Rubtsova – Investigation (XRD analysis), N.Yu. Tabachkova – Investigation (TEM), A.V. Nokhrin & A.I. Malkin – Formal analysis, Writing - original draft preparation, Data curation, Visualization; A.M. Bakhmetyev, P.V. Tryaev, R.A. Vlasov – Investigation (HSC test).

**Funding:** The work was carried out within the framework of the grant #H-498-99_2021-2023 (#075-15-2021-1332) of the Federal Academic Leadership Priority Program Priority-2030 of the Ministry of Science and Higher Education of the Russian Federation and the grant #075-03-2023-096 (FSWR-2023-0037) of the Ministry of Science and Higher Education of the Russian Federation. The investigations with the use of TEM were carried out at the Center Collective Use "Material Science and Metallurgy" of the National University of Science and Technology "MISIS" (grant #075-15-2021-696 of the Ministry of Science and Higher Education of the Russian Federation).

**Conflicts of Interest:** The authors declare that they have no known competing financial interests or personal relationships that could have appeared to influence the work reported in this paper.

**References**

1. I.V. Gorynin, B.B. Chechulin, Titan in Mechanical Engineering, Mashinostroenie, Moscow, Russia, 1990. (In Russian)
2. A.S. Oryshchenko, I.V. Gorynin, V.P. Leonov, A.S. Kudryavtsev, V.I. Mikhaylov, E.V. Chudakov, Marine titanium alloys: Present and future, Inorg. Mater. Appl. Res. 6 (2015) 571–579. https://doi.org/10.1134/S2075113315060106
3. I.V. Gorynin, Titanium alloys for marine application, Mater. Sci. Eng. A. 263 (1999) 112–116. https://doi.org/10.1016/S0921-5093(98)01180-0
4. E.A. Saunders, T.P. Chapman, A.R.M. Walker, et al., Understanding the "blue spot": Sodium chloride hot salt stress-corrosion cracking in titanium-6246 during fatigue testing at low pressure, Engineering Failure Analysis. 61 (2016) 2-20. https://doi.org/10.1016/j.engfailanal.2015.06.008

5. L. Fan, L. Liu, Y. Cui, et al., Effect of streaming water vapor on the corrosion behavior of Ti60 alloy under solid NaCl deposit in water vapor at 600 ºC, Corrosion Science. 160 (2019) 108177. https://doi.org/10.1016/j.corsci.2019.108177

6. Chevrot, T. Pressure Effects on the Holt-Salt Stress-Corrosion Cracking of Titanium Alloys. Ph.D. Thesis, Cranfield University, Bedford, UK, 1994.

7. S. Joseph, T.C. Lindley, D. Dye, E.A. Saunders, The mechanisms of hot salt stress corrosion cracking in titanium alloy Ti-6Al-2Sn-4Zr-6Mo, Corrosion Science. 134 (2018) 169-178. https://doi.org/10.1016/j.corsci.2018.02.025

8. D. Sinigaglia, G. Taccani, B. Vicentini, Hot-salt-stress-corrosion cracking of titanium alloys. Corrosion Science. 18 (1978) 781-796. https://doi.org/10.1016/0010-938X(78)90015-X

9. J.R. Meyers, J.A. Hall, Hot salt stress corrosion cracking of titanium alloys: An improved model for the mechanism, Corrosion, 33 (1977) 252-257. https://doi.org/10.5006/0010-9312-33.7.252

10. V.C. Peterson, Hot-salt stress-corrosion o titanium: A review of the problem and methods for improving the resistance of titanium, JOM. 23 (1971) 40-47. https://doi.org/10.1007/BF03355696

11. Y. Shi, S. Joseph, E.A. Saunders, et al., AgCl-induced hot salt stress corrosion cracking in a titanium alloy, Corrosion Science. 187 (2021) 109497. https://doi.org/10.1016/j.corsci.2021.109497

12. B.A. Kolachev, V.V. Travkin, Yu.N. Artsybasov, Effect of salt corrosion on the endurance strength of pseudoalfa-titanium alloys, Materials Science. 13 (1978) 391-394. https://doi.org/10.1007/BF00715257

13. V.V. Travkin, V.F. Pshirkov, B.A. Kolachev, Thermodynamic analysis of the chemical mechanism for the hot-salt corrosion of titanium alloys, Materials Science. 15 (1979) 134-137. https://doi.org/10.1007/BF00716248

14. M.D. Pustode, V.S. Raja, N. Paulose, The stress-corrosion cracking susceptibility of near-α titanium alloy IMI 834 in presence of hot salt, Corrosion Science. 82 (2014) 191-196. https://doi.org/10.1016/j.corsci.2014.01.013

15. T.P. Chapman, R.J. Chater, E.A. Saunders, et al., Environmentally assisted fatigue crack nucleation in Ti-6Al-2Sn-4Zr-6Mo, Corrosion Science. 96 (2015) 87-101. https://doi.org/10.1016/j.corsci.2015.03.013

16. W. Chen, W. Wang, L. Liu, et al., The effect of NaCl-induced corrosion on Ti60's hot salt stress corrosion cracking, Corrosion Science. 226 (2024) 111677. https://doi.org/10.1016/j.corsci.2023.111677

17. W. Chen, R. Li, L. Liu, et al. Effect of NaCl-rich environment on internal corrosion for Ti60 alloy at 600 ºC, Corrosion Science. 220 (2018) 111307. https://doi.org/10.1016/j.corsci.2023.111307

18. V.A. Kolachev, V.V. Travkin, The influence of alloying on the tendency of titanium alloys towards salt corrosion, Materials Science. 17 (1980) 136-140. https://doi.org/10.1007/BF00722899

19. V.A. Kolachev, V.V. Travkin, Effect of aluminium on the susceptibility of titanium to corrosion by salt, Materials Science. 12 (1977) 630-633. https://doi.org/10.1007/BF00721765

20. C. Ciszak, I. Abdallah, I. Popa, et al., Degradation mechanism of Ti-6Al-2Sn-4Zr-2Mo-Si alloy exposed to sold NaCl deposit at high temperature, Corrosion Science. 172 (2020) 108611. https://doi.org/10.1016/j.corsci.2020.108611

21. V.P. Batrakov, L.N. Pivovarova, L.V. Zakharova, Salt corrosion of titanium alloys, Metal Science and Heat Treatment. 16 (1974) 418-419. https://doi.org/10.1007/BF00652203

22. M. Li, D. Liu, Z. Bai, et al., Hot salt corrosion behavior of Ti-6Al-4V alloy treated with different surface deformation strengthening processes, Materials Chemistry and Physics. 309 (2023) 128410. https://doi.org/10.1016/j.matchemphys.2023.128410

23. H. Shi, D. Liu, T. Jia, et al., Effect of the ultrasonic surface rolling process and plasma electrolytic oxidation on the hot salt corrosion fatigue behavior of TC11 alloy, International Journal of Fatigue. 168 (2023) 107443. https://doi.org/10.1016/j.ijfatigue.2022.107443

24. Y. Hua, Y. Bai, Y. Ye, et al., Hot corrosion behavior of TC11 titanium alloy treated by laser shock processing, Applied Surface Science. 283 (2013) 775-780. https://doi.org/10.1016/j.apsusc.2013.07.017

25. V. Chakkravarthy, J.P. Oliveira, A. Mahomed, et al., Effect of abrasive water jet peening of NaCl-induced hot corrosion behavior of Ti-6Al-4V, Vacuum. 210 (2023) 111872. https://doi.org/10.1016/j.vacuum.2023.111872

26. S. Kumar, K. Chattopadhyay, G.S. Mahobia, V. Singh, Hot corrosion behaviour of Ti-6Al-4V modified by ultrasonic shot peening, Materials and Design. 110 (2016) 196-206. https://doi.org/10.1016/j.matdes.2016.07.133

27. M. Encrenaz, P. Faure, J.A. Petit, Hot salt stress corrosion resistance of Ti 6246 alloy, Corrosion Science. 40 (1998) 939-950. https://doi.org/10.1016/S0010-938X(98)00028-6

28. C. Ciszaka, I. Popa, J.-M. Brossard, et al., NaCl induced corrosion of Ti-6Al-4V alloy at high temperature. Corrosion Science, 110 (2016) 91–104. https://doi.org/10.1016/j.corsci.2016.04.016

29. M.D. Pustode, V.S. Raja, Hot salt stress corrosion cracking behavior of Ti-6242S alloy, Metallurgical and Materials Transactions A. 46 (2015) 6081–6089. https://doi.org/10.1007/s11661-015-3155-2

30. H. Shi, D. Liu, X. Zhang, et al., Effect of pre-hot salt corrosion on hot salt corrosion fatigue behavior of the TC11 titanium alloy at 500 ºC, International Journal of Fatigue. 163 (2022) 107055. https://doi.org/10.1016/j.ijfatigue.2022.107055

31. R.-Z. Li, J.-B. Pu, C.-Q. Cheng, et al., Effect of hot corrosion on cycle deformation and fracture behavior of Ti-6Al-4V alloy under salt coating, Corrosion Science. 224 (2023) 111545. https://doi.org/10.1016/j.corsci.2023.111545

32. M.D. Pustode, V.S. Raja, B. Dewangan, N. Paulose, Effect of long term exposure and hydrogen effects on HSSCC behavior of titanium alloy IMI 834, Materials and Design. 86 (2015) 841-847. https://doi.org/10.1016/j.matdes.2015.08.002

33. S. Joseph, P. Kontis, Y. Chang, et al., A cracking oxygen story: A new view of stress corrosion cracking in titanium alloys, Acta Materialia. 227 (2022) 117687. https://doi.org/10.1016/j.actamat.2022.117687

34. S. Li, X. Xu, X. Bai, B. Cao, Hot corrosion resistance of TB8 titanium alloy after ECAP and heat treatment, Journal of Wuhan University of Technology. 38 (2023) 1440-1448. http://dx.doi.org/10.1007/s11595-023-2840-z

35. V.N. Chuvil'deev, V.I. Kopylov, A.V. Nokhrin, et al., Study of mechanical properties and corrosive resistance of ultrafine-grained α-titanium alloy Ti-5Al-2V, Journal of Alloys and Compounds. 723 (2017) 354-367. https://doi.org/10.1016/j.jallcom.2017.06.220

36. V.N. Chuvil'deev, V.I. Kopylov, A.V. Nokhrin, et al., Effect of severe plastic deformation realized by rotary swaging on the mechanical properties and corrosion resistance of near-α-titanium alloy Ti-2.5Al-2.6Zr, Journal of Alloys and Compounds. 785 (2019) 1233-1244. http://doi.org/10.1016/j.jallcom.2019.01.268

37. V.N. Chuvil'deev, V.I. Kopylov, A.V. Nokhrin, et al., Corrosion fatigue crack initiation in ultrafine-grained near-α titanium alloy PT7M prepared by Rotary Swaging, Journal of Alloys and Compounds. 790 (2019) 347-362. http://doi.org/10.1016/j.jallcom.2019.03.146

38. R.Z. Valiev, T.G. Langdon, Principles of equal-channel angular pressing as a processing tool for grain refinement, Progress in Materials Science. 51 (2006) 881-981. https://doi.org/10.1016/j.pmatsci.2006.02.003

39. A.Yu. Vinogradov, V.V. Stolyarov, S. Hashimoto, R.Z. Valiev, Cyclic behavior of ultrafine-grain titanium produced by severe plastic deformation, Materials Science and Engineering A. 318 (2001) 163-173. https://doi.org/10.1016/S0921-5093(01)01262-X

40. S. Zherebtsov, E. Kudryavtsev, S. Kostjuchenko, et al., Strength and ductility-related properties of ultrafine grained two-phase titanium alloy produced by warm multiaxial forging, Materials Science and Engineering A. 536 (2012) 190-196. https://doi.org/10.1016/j.msea.2011.12.102

41. Y. Gu, J. Jiang, A. Ma, et al., Enhanced corrosion behavior of ultrafine-grained pure titanium in simulated high-temperature seawater, Journal of Materials Research and Technology. 19 (2022) 1-6. https://doi.org/10.1016/j.jmrt.2022.05.032

42. Y. Gu, J. Jiang, A. Ma, et al., Corrosion behavior of pure titanium processed by rotary-die ECAP, Journal of Materials Research and Technology. 15 (2021) 1873-1880. https://doi.org/10.1016/j.jmrt.2021.09.047

43. S. Huang, Y. Jin, Y. Wang, et al., Stress corrosion cracking of ultrafine-grained Ti-2Fe-0.1B alloying after equal channel angular pressing, Metals. 13 (2023) 1316. https://doi.org/10.3390/met13071316

44. I. Ratochka, O Lykova, I. Mishin, E. Naydenkin, Superplastic deformation behavior of Ti-4Al-2V alloy governed by its structure and precipitation phase evolution, Materials Science and Engineering A. 731 (2018) 577-582. https://doi.org/10.1016/j.msea.2018.06.094

45. A.O. Mosleh, A.V. Mikhaylovskaya, A.D. Kotov, et al., Superplastic deformation behavior of ultra-fine-grained Ti-1V-4Al-3Mo alloy: constitutive modeling and processing map, Materials Research Express. 6 (2019) 096584. https://doi.org/10.1088/2053-1591/ab31f9

46. I.V. Ratochka, O.N. Lykova, E.V. Naidenkin, Influence of low-temperature annealing time on the evolution of the structure and mechanical properties of a titanium Ti-Al-V alloy in the submicrocrystalline state, The Physics o Metals and Metallography. 116 (2015) 302-308. https://doi.org/10.1134/S0031918X15030114

47. V.N. Chuvil'deev, A.V. Nokhrin, V.I. Kopylov, et al., Spark Plasma Sintering for high-speed diffusion bonding of the ultrafine-grained near-α Ti-5Al-2V alloy with high strength and corrosion resistance for nuclear engineering, Journal of Materials Science. 54 (2019) 14926-14949. https://doi.org/10.1007/s10853-019-03926-6

48. V. Chuvil'deev, A. Nokhrin, C. Likhnitskii, et al., Corrosion resistance of the weld joints from the ultrafine-grained near-α titanium alloys Ti-5Al-2V obtained by Spark Plasma Sintering, Metals. 13 (2023) 766. https://doi.org/10.3390/met13040766

49. Q. Mao, Y. Li, Y. Zhao, A review of mechanical properties and microstructure of ultrafine grained metals and alloys processed by rotary swaging, Journal of Alloys and Compounds. 896 (2022) 163122. https://doi.org/10.1016/j.jallcom.2021.163122

50. Q. Mao, X. Chen, J. Li, Y. Zhao, Nano-gradient materials prepared by rotary swaging, Nanomaterials. 11 (2021) 2223. https://doi.org/10.3390/nano11092223

51. E.V. Naydenkin, I.P. Mishin, O.V. Zabudchenko, et al., Structural-phase state and mechanical properties of β titanium alloy produced by rotary swaging with subsequent aging, Journal of Alloys and Compounds. 935 (2023) 167973. https://doi.org/10.1016/j.jallcom.2022.167973

52. L. Kunčická, Strucutral phenomena introduced by Rotary Swaging: A review, Materials. 17 (2024) 466. https://doi.org/10.3390/ma17020466

53. B.S. Rodchenkov, A.V. Kozlov, Yu.G. Kuznetsov, et al., Irradiation behavior of Ti-4Al-2V (ПТ-3В) alloy for ITER blanket modules flexible attachment, Journal of Nuclear Materials. 367-370 (2007) 1312-1315. https://doi.org/10.1016/j.jnucmat.2007.03.261

54. V.I. Kopylov, A.V. Nokhrin, N.A. Kozlova, et al., Effect of σ-phase on the strength, stress relaxation behavior, and corrosion resistance of an ultrafine-grained austenitic steel AISI 321, Metals. 13 (2023) 45. https://doi.org/10.3390/met13010045

55. O.H. Basquin, The exponential law of endurance tests, Proc. ASTM. (1910) 625–630.

56. A.M. Bakhmetyev, O.A. Bykh, N.G. Sandler, et al., Hot-salt corrosion of alloys PT-7M, 42ХНМ, Inconel 690 and Incoloy 800, Applied Solid State Chemistry. 1 (2019) 23–31. https://doi.org/10.18572/2619-0141-2019-1-23-31

57. I.V. Pyshmintsev, Y.I. Kosmatkii, E.A. Gornostaeva, et al., Structure, phase composition and mechanical properties of hot-extruded Ti-3Al-2.5V pipe after vacuum annealing, Metallurgist. 63 (2019) 751-758. https://doi.org/10.1007/s11015-019-00885-w

58. L. Yuan, W. Wang, H. Zhang, et al., Evolution of texture and tensile properties of cold-rotary swaged Ti-6Al-4V alloy seamless tubing after annealing at different temperatures, Materials Science and Engineering A. 841 (2022) 143003. https://doi.org/10.1016/j.msea.2022.143003

59. J.A. Fellows, Fractography and Atlas of Fractographs, Metals Handbook, Metals Park, Ohio: Am. Soc. Met., 1974

60. V.M., Segal, I.J. Beyerlein, C.N. Tome, V.N. Chuvil'deev, V.I. Kopylov, Fundamentals and Engineering of Severe Plastic Deformation, Nova Science Publishers, New York, 2010.

61. A.V. Semenycheva, V.N. Chuvil'deev, A.V. Nokhrin, A theoretical model of grain boundary self-diffusion in metals with phase transitions (case study into titanium and zirconium), Physics B: Condensed Matter. 537 (2018) 105-110. https://doi.org/10.1016/j.physb.2018.01.069

62. G. Lütjering, J.C. Williams, Titanium, 2nd ed.; Springer: Berlin/Heidelberg, Germany, 2007

63. A.A. Murashov, N.N. Berendeev, A.V. Nokhrin, et al., Investigation of the processes of fatigue and corrosion-fatigue destruction of pseudo-α titanium alloy, Inorganic Materials: Applied Research. 13 (2022) 349-356. http://dx.doi.org/10.1134/S2075113322020290

64. J. Yang, Y. Song, K. Dong, E-H. Han, Research progress on the corrosion behavior of titanium alloys, Corrosion Reviews. 41 (2023) 5-20. https://doi.org/10.1515/corrrev-2022-0031